\documentclass[twocolumn,superscriptaddress,showpacs,preprintnumbers,amsmath,amssymb,prb]{revtex4}

\usepackage{graphicx}
\usepackage{dcolumn}
\usepackage{bm}
\usepackage{verbatim}
\usepackage{multirow}

\newcommand{\Asub}{A_{\text{sub}}}
\newcommand{\Afilm}{A_{\text{film}}}
\newcommand{\Amol}{A_{\text{mol}}}
\newcommand{\phisub}{\phi_{\text{sub}}}
\newcommand{\supth}{^{\text{th}}}
\newcommand{\parcfrac}{\delta}
\newcommand{\sflow}{\text{SF}}
\newcommand{\vic}{\text{vic}}
\newcommand{\ter}{\text{ter}}
\newcommand{\lbl}{\text{LBL}}
\newcommand{\etal}{\textit{et al.}}
\newcommand{\insitu}{\textit{in situ}}
\newcommand{\Insitu}{\textit{In situ}}
\begin{document}

\preprint{DRAFT 15 -- With b/w compatible version of Fig. 1}

\title{Quantitative modeling of \textit{in situ} x-ray reflectivity during organic molecule thin film growth}

\author{Arthur R. Woll}
\affiliation{Cornell High Energy Synchrotron Source, Cornell University, Ithaca, NY 14853, USA}
\author{Tushar V. Desai}
\affiliation{School of Chemical and Biomolecular Engineering, Cornell University, Ithaca, NY 14853, USA}
\author{James R. Engstrom}
\affiliation{School of Chemical and Biomolecular Engineering, Cornell University, Ithaca, NY 14853, USA}

\date{\today}

\begin{abstract}
Synchrotron-based x-ray reflectivity is increasingly employed as an \textit{in situ} probe of surface morphology during thin film growth, but complete interpretation of the results requires modeling the growth process. Many models have been developed and employed for this purpose, yet no detailed, comparative studies of their scope and accuracy exists in the literature. Using experimental data obtained from hyperthermal deposition of pentane and diindenoperylene (DIP) on SiO$_2$, we compare and contrast three such models, both with each other and with detailed characterization of the surface morphology using ex-situ atomic force microscopy (AFM). These two systems each exhibit particular phenomena of broader interest: pentacene/SiO$_2$ exhibits a rapid transition from rough to smooth growth. DIP/SiO$_2$, under the conditions employed here, exhibits growth rate acceleration due to a different sticking probability between the substrate and film. In general, \textit{independent of which model is used}, we find good agreement between the surface morphology obtained from fits to the \insitu x-ray data with the actual morphology at early times. This agreement deteriorates at later time, once the root-mean squared (rms) film roughness exceeds about 1 ML.  A second observation is that, because layer coverages are under-determined by the evolution of a single point on the reflectivity curve, we find that the best fits to reflectivity data  \---- corresponding to the lowest values of $\chi_\nu^2$ \---- do not necessarily yield the best agreement between simulated and measured surface morphologies. Instead, it appears critical that the model reproduce all local extrema in the data. In addition to showing that layer morphologies can be extracted from a minimal set of data, the methodology established here provides a basis for improving models of multilayer growth by comparison to real systems.
\end{abstract}


\maketitle

\section{Introduction} \label{intro}

\Insitu, surface-sensitive scattering techniques, such as reflection high energy electron diffraction (RHEED) and x-ray scattering at the so-called ``anti-Bragg'' position\cite{Neave:1985ga,Cohen:1986gz,Fuoss:1992aw}, yield direct information about surface morphology during growth, and have been applied to virtually all methods of thin film growth\---such as electrodeposition\cite{Magnussen:2008ek}, molecular beam epitaxy (MBE), chemical vapor deposition (CVD)\cite{Murty:1999is,Stephenson:1999dz}, and pulsed laser deposition (PLD)\cite{Lippmaa:2000kg,Eres:2002zy,Fleet:2005xy,Fleet:2006lv,Tischler:2006yv,Dale:2006xw}. Their principal advantages, compared to scan-probe microscopies such as atomic force microscopy (AFM), are their time resolution and the ease with which they can be incorporated into growth systems as qualitative, \textit{in situ} probes. For example in homoepitaxy, growth is classified as ``step-flow''(SF), ``layer-by-layer''(LBL) or ``three-dimensional'' based, respectively, on whether the specular reflectivity remains constant, oscillates, or decreases monotonically with increasing film thickness. An additional advantage of x-ray scattering is the accuracy with which precise calculations of scattered intensity can be performed. Increasingly, this has been exploited\cite{Vandervegt:1992va,Woll:1999at,Braun:2003ld, Mayer:2004hs, Kowarik:2007kf, Kowarik:2009ip, Amassian:2009mk, Amassian:2009yk, Desai:2010pfp} to extract quantitative information about the surface morphology, namely the layer coverages as a function of time $\theta_n(t)$, during growth. This analysis, however, requires a specific model of the morphological evolution of the film. 

\begin{figure}
\includegraphics[scale=1]{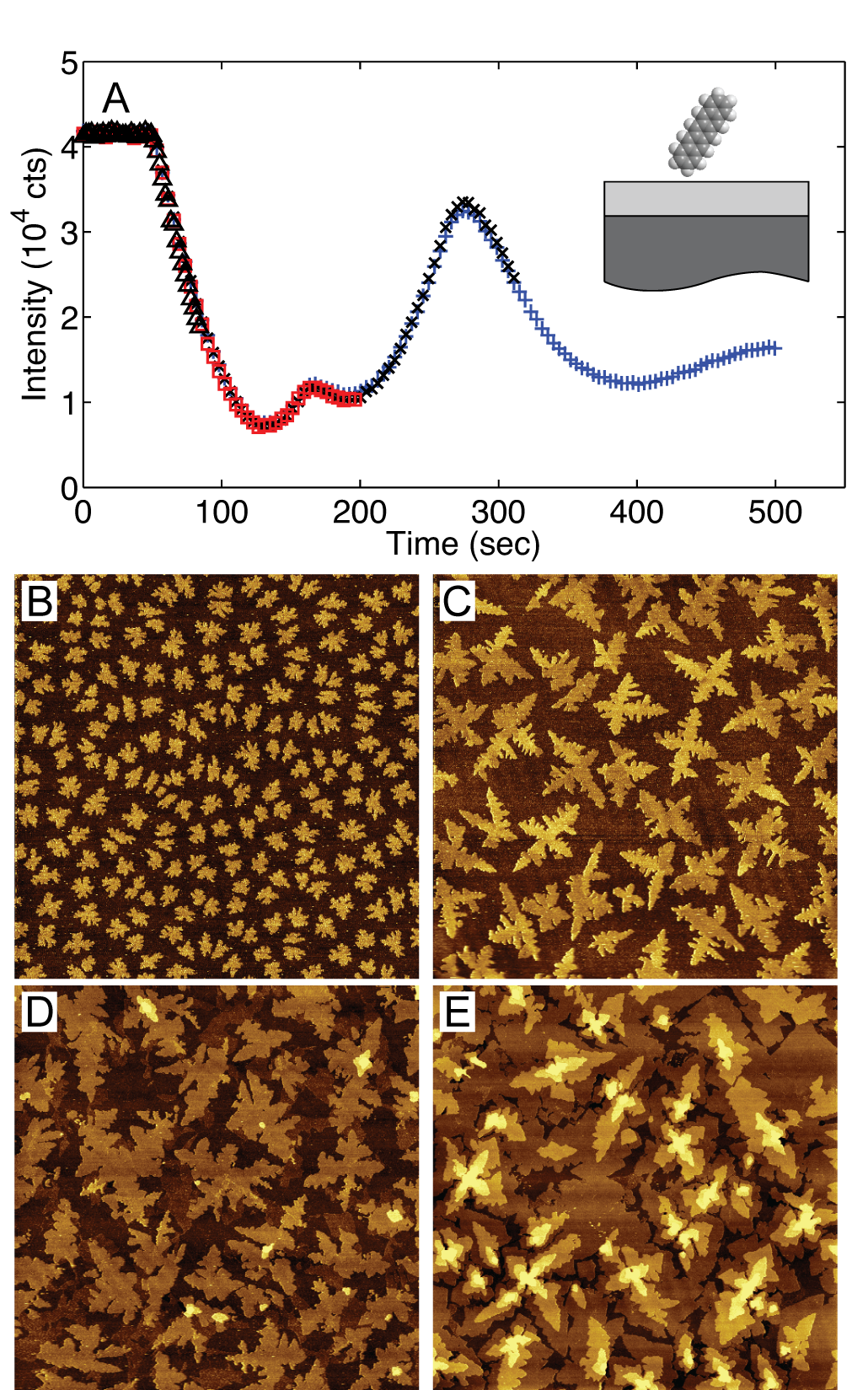}
\caption{\label{fig:pent_raw} (Color online) (A) XRR data obtained near the pentacene anti-Bragg position during pentacene deposition on SiO$_2$. Data are shown for four separate films grown to thickness of ($\triangle$) 0.34 ML, ($\square$) 1.4 ML,  ($\times$) 2.5 ML , and ($+$) 4.4 ML, performed in immediate succession on the same substrate under the same conditions. (B-E) AFM images of the four films represented in (A) in order of increasing thickness. All images are 20 $\mu$m$\times$20$\mu$m.}
\end{figure}

Figure \ref{fig:pent_raw} shows x-ray reflectivity (XRR) and AFM data obtained from four pentacene films, grown in immediate succession on a single SiO$_2$ substrate under nominally identical conditions (substrate temperature 48$^\circ$C), and serves to illustrate the main subject of this work: how accurately can the height probability distribution, which may be obtained directly from the AFM data in Fig.~\ref{fig:pent_raw}B-E, be extracted from fits to the x-ray data in Fig.~\ref{fig:pent_raw}A? In addition, how does the choice of model used for such analysis affect the simulated morphology?

The models described here are necessary, in part, because x-ray intensity measurements such as those in Figure~\ref{fig:pent_raw}A are performed at only one point in reciprocal space. As has recently been demonstrated\cite{Braun:2003ld,Kowarik:2007kf, Kowarik:2008hx,Kowarik:2009ie,Kowarik:2009ip}, measuring multiple points along the reflectivity curve can reduce or potentially obviate the requirement for such models. However, such measurements are not always possible or compatible with a particular system or experiment. Further, with notable exceptions\cite{Kowarik:2009ip}, such enhanced measurements are often used not to replace the use of growth models, but to limit uncertainties associated with comparing measurements to calculations. Finally, although this manuscript deals with the limited problem of determining the surface morphology of a growing film, the models described here also concern the broader challenge of understanding, in detail, how microscopic, kinetic parameters of a system determine its morphology. Clearly, comparing the simulated morphologies of these models to real systems is a critical, first step in validating them for deeper, more general use.

One motivation for this work is to evaluate a particular model (the ``modified Cohen'' model, see section~\ref{sect:models} below) we recently developed to measure the time-dependent growth rate of hyperthermally-deposited organic thin films\cite{Amassian:2009mk}. This time-dependence arises from differences between the sticking probabilities of deposited molecules on the bare substrate and the growing film. Because this sticking probability depends strongly on the substrate, a quartz crystal monitor is an inappropriate measure of growth rate. This phenomena has been observed in other systems: in metal-organic based GaN growth on sapphire, both growth rate acceleration\cite{Woll:1999at} and deceleration\cite{Headrick:1998kw} have been observed using \insitu measurements of Ga K$\alpha$ x-ray fluorescence. For organic molecules containing only light elements, this method is not possible. Growth rate acceleration may also be determined using ex situ microscopy (see section \ref{sect:dip} below), but this approach requires several samples per growth condition, and can be severely complicated by the presence of de-wetting, a common phenomena in organic thin films\cite{Amassian:2009yk}. Accurate determination of growth rates and growth rate acceleration from time-resolved XRR data relies directly on quantitative analysis as described above, and thus on the accuracy of the growth simulation. In prior studies, we validated our model by comparing both the total film thickness and rms roughnesses of thin films measured by AFM to those predicted by best-fit simulations. Here, we provide a complete description of this model, and also address the specific problem of measuring growth rate acceleration. 

We proceed, in section~\ref{sect:xray}, by reviewing XRR intensity calculations relevant for thin film growth. To elucidate how different features of the film and substrate influence the character of oscillations, we explicitly derive x-ray scattering parameters in the approximation where each layer consists of uniform density slabs. Additional insight is gained by considering the case of continuous, or step-flow growth, although it is not directly relevant for the particular systems described here. In section~\ref{sect:models}, we provide detailed descriptions of three distinct deterministic, rate-equation models of thin film growth. Finally, in section~\ref{sect:results}, we compare these models to two particular systems. The first of these, pentacene/SiO$_2$ (represented in Fig.~\ref{fig:pent_raw}), has well-known structural and morphological evolution and, under these growth conditions (see section \ref{sect:pent}), does not exhibit growth rate acceleration. Four films of varying film thickness are grown under identical conditions. Height distributions of these films, measured with ex situ AFM, are then compared in detail with those deduced using fits to reflectivity data at the anti-Bragg position. The second system described in this paper, diindenoperylene (DIP)/SiO$_2$, exhibits growth rate acceleration, which we model as a difference in molecular sticking probability between the substrate and deposited film. We find that two of the models, including that in Ref.~\onlinecite{Amassian:2009mk}, accurately describe the degree of acceleration. 

\section{Experimental} \label{sect:experiment}

Experiments were conducted at the G3 station at the Cornell Higher Energy Synchrotron Source (CHESS) in a custom-designed, ultrahigh vacuum chamber with a base pressure of ~5$\times$10$^{-9}$ mbar described previously \cite{Schroeder:2004phd,Amassian:2009mk}. Growth was performed on 300 nm-thick SiO$_2$ thermal oxide films on Si(001) wafers, using a hyperthermal molecular source. The deposition energy, measured using time-of-flight mass spectrometry, was approximately 2.5 eV for pentacene and 4.2 eV for DIP.  Synthetic W/B$_4$C multilayer monochromators set the beam energy to 9.75~keV with $\Delta E/E\approx1.5$\%.  The unattenuated flux at the sample position was approximately 5x10$^{13}$ photons\--s$^{-1}$\--mm$^{-2}$. In order to avoid radiation damage, the beam was attenuated by a factor of 10 or more using multiple layers of aluminum foil at the upstream end of the hutch. An avalanche photodiode detector (Oxford Danfysik, Oxford, UK) scintillator counter was used for measuring the scattered x-ray intensity. AFM was conducted ex situ in tapping mode using a Digital Instruments 3100 Dimension microscope (Santa Barbara, CA). 

\section{Theory} \label{sect:theory}

\subsection{X-ray Scattering} \label{sect:xray}

The computational simplicity of x-ray scattering results from the fact that, far from a Bragg peak or from the critical angle for total external reflection, the Born approximation (also referred to as the single scattering or kinematic approximation) may be used \cite{Sinha:1988qb}. In this approximation, the scattered intensity $I(\vec{q})$ is proportional to $|A(\vec{q})|^2$,  where the scattering amplitude $A(\vec{q})$ is the Fourier transform of the electron density:
\begin{equation}
  \label{eq:xray1}
  A(\vec{q}) =\int_V \rho(\vec{r})\exp(-i \vec{q} \cdot \vec{r}) \text{.} 
\end{equation}
For the particular case of \insitu data collected during thin film growth at a position $\vec{q} = q_z\hat{z}$ in reciprocal space, and with the assumption that the substrate does not vary during deposition, this may be simplified to ~\cite{Woll:1999at,Mayer:2004hs,Kowarik:2009ie},
\begin{equation}
  \label{eq:xray2}
A(q_z, t)  = \Asub e^{i\phisub} + \Afilm\sum_{n=1} \theta_n(t) e^{-iq_zc(n-1)}   \text{.} 
\end{equation}
Here, $\theta_n$ is the fractional coverage of layer $n$, and $c$ is the lattice parameter of the film normal to the surface. Typically, experiments are conducted at the anti-Bragg position, corresponding to $q_z c = \pi$. Since $\Afilm$ and $\Asub$ are both defined as pure real, there are only two unknown scattering parameters, $\phisub$ and the ratio $\Afilm/\Asub$. These parameters may be calculated directly if the atomic structure\---including the detailed structure of the interface\---is known. For crystalline substrates, $\Asub e^{i\phisub}$ incorporates a semi-infinite sum over the buried substrate lattice. Krause \etal\cite{Krause:2004dk,Krause:2004ss} explicitly calculated $A(q_z, t) $ for the case of the organic molecule growth on Ag(111). Alternatively, they can be treated as free parameters.

\begin{figure}
\includegraphics[scale=0.5]{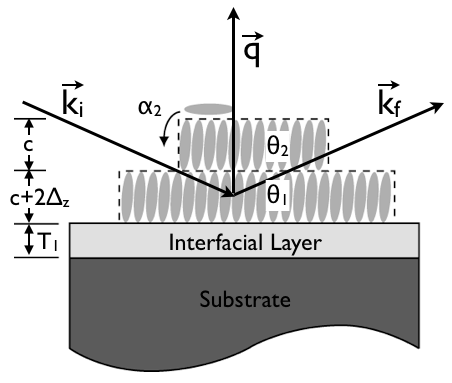}
\caption{\label{fig:real_schem} Real-space schematic representation of organic thin film growth. Parameters $c$, $\tau$, $\Delta$, $\theta_n$, and $\alpha_n$ are described in the text. $\vec{k_i}$,$\vec{k_f}$, and $\vec{q}$ correspond to the incident and final wave vectors, and momentum transfer, respectively, in specular geometry.}
\end{figure}

In appendix \ref{appendixa} we derive a general expression for $\Asub$ and $\phisub$ under the assumption that the sample geometry resembles that of Fig. \ref{fig:real_schem}. 
Specifically, we assume that the substrate is composed of a thin amorphous layer with density $\rho_1$ and thickness $T_1=c\tau$ on an amorphous, semi-infinite layer with density $\rho_0$. This model is particularly suitable for amorphous substrates such as SiO$_2$, but also valid for Si substrates with a thin SiO$_2$ native oxide\cite{Ruiz:2003um}, and thick, thermal SiO$_2$ layers covered by a self-assembled monolayer\cite{Mayer:2006da,Amassian:2009yk} or an interfacial water layer \cite{Mayer:2004hs,Wo:2006nc}. 

To help visualize the scattering intensity arising from Eq.\ref{eq:xray2} near the anti-Bragg position for organic thin films, it is conceptually useful to further approximate the film as composed of uniform density slabs of height $c$ and density $\rho_2$. This approximation explicitly eliminates Bragg Peaks, but is nevertheless reasonably accurate near the anti-Bragg position whenever this geometry probes a length scale, $l=2\pi/q_z=2c$, that is large compared to interatomic distances. For example, for pentacene and DIP, $c\approx$1.54 nm and 1.66 nm\cite{Kowarik:2009og}, respectively.  In this approximation, we have the result
\begin{equation}
\label{eq:asub}
A_{\text{sub}} e^{i\phisub} =  \frac{c }{2\pi L} i e^{i \pi L (1+2\Delta)}(\rho_1 + (\rho_0-\rho_1)e^{i 2\pi L\tau} ) 
\text{.}
\end{equation}
$\phisub$ is simply the complex phase of $A_{\text{sub}}$ above, while $\Afilm/\Asub$ is 
\begin{equation}
\label{eq:pfilm}
\frac{\Afilm}{\Asub}=\frac{2\rho_2 \sin{\pi L}}{\left | \rho_1 + (\rho_0-\rho_1)e^{i 2\pi L\tau} \right |}
\text{.}
\end{equation}
In Eqs.~\ref{eq:asub} and \ref{eq:pfilm}, $L = q_z/(2\pi/c)$, and $\Delta$ corresponds to a small difference, as a fraction of unit cell height $c$, in height of the center of the first deposited layer from $c/2$ above the substrate surface. Despite the fact that these equations include more than the two original parameters in Eq.~\ref{eq:xray2}, this representation generally reduces parameter space, since these parameters are at least approximately known, and can thus be constrained.

\begin{figure}
\includegraphics[scale=1]{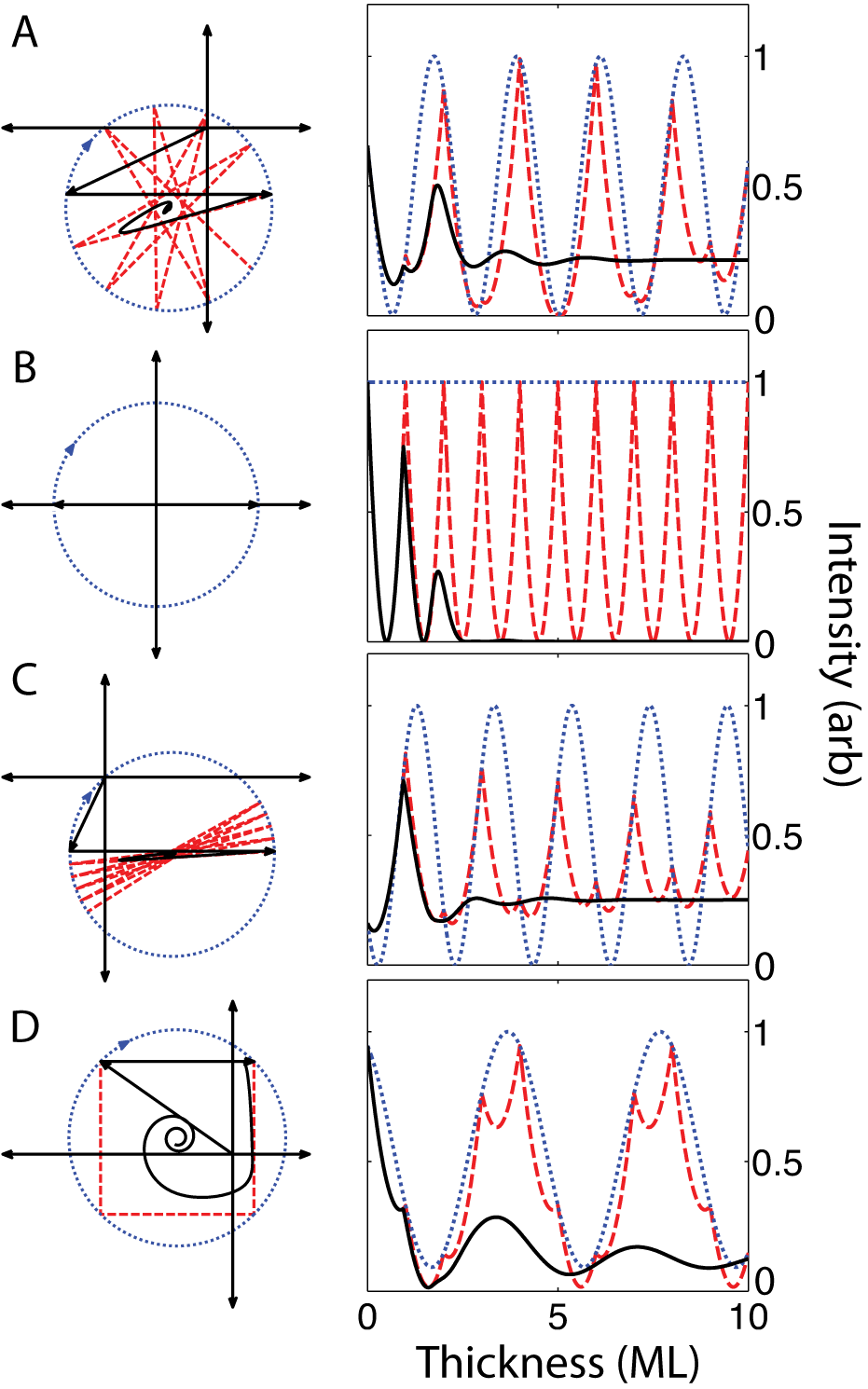}
\caption{\label{fig:xray_schem} (Color online) (left) Complex scattering amplitude and (right) intensity under identical growth conditions but different scattering conditions. In all plots, the dashed and dotted curves (color online) correspond to ideal LBL and SF growth, respectively, while the solid curve corresponds to growth according to the best-fit curve of the data in Fig.~\ref{fig:pent_raw}A. The four pairs of plots simulate (A) pentacene/SiO$_2$ identical to that found from the fit to the modified Cohen model shown in Fig.~\ref{fig:pent_xray}; (B) homoepitaxy at the anti-Bragg position (L=0.5); (C) pentacene on a 1 nm water on SiO$_2$; (D) pentacene/SiO$_2$ measured at the quarter-Bragg position (L=0.25).}
\end{figure}

Recently, Kowarik \etal\cite{Kowarik:2009ie} described the varying appearance of anti-Bragg intensity oscillations caused by variations in the two parameters $\phisub$ and $\Afilm/\Asub$. Equations~\ref{eq:asub} and \ref{eq:pfilm} connect these parameters with details of the system. To help interpret and illustrate Eq.~\ref{eq:xray2}, the evolution of both the scattering amplitude and intensity under four distinct scattering conditions are shown in Fig.~\ref{fig:xray_schem}. In all eight plots, the solid curve corresponds to the same evolution of coverages $\theta_n(t)$: namely, those obtained from a fit of the ``modified Cohen'' model (see section~\ref{sect:models}) to the data in Fig.~\ref{fig:pent_raw}A. The dashed curves correspond to ideal LBL growth, while the dotted curves correspond to ideal SF growth. Although SF growth is not relevant for the examples discussed below, we nevertheless plot it to explicitly demonstrate the relationship between ``roughness'' and ``Kiessig'' oscillations. The scattering amplitude during SF growth is derived in appendix \ref{appendixb}. 

In each of the left-hand plots in Fig.~\ref{fig:xray_schem}, $\Asub e^{i\phisub}$ is the vector originating from the origin, while $\Afilm$ originates at the tip of $\Asub e^{i\phisub}$ and extends along the positive real axis. Figure~\ref{fig:xray_schem}A corresponds to the precise parameters obtained from the fit, as described below and shown in Table~\ref{tab:pent_pars}. Because the best-fit value of $L$ is somewhat less than 0.5, the scattering amplitude of ideal LBL growth traverses a non-convex regular polygon. The subsequent pairs of plots simulate other cases of interest. For example, Fig.~\ref{fig:xray_schem}B corresponds to the familiar case of homoepitaxy, which exhibits two one oscillation per layer in LBL growth, and constant intensity during SF growth.  Figure~\ref{fig:xray_schem}C corresponds to growth on a ~1 nm water layer on SiO$_2$. In this case, the SiO$_2$ substrate and water layer nearly cancel ($\rho_0/\rho_1 \approx 2.2\Rightarrow\rho_1 \approx \rho_0-\rho_1$, and $\exp(i2\pi L \tau)\approx-1$), so that the ratio $\Afilm/\Asub$ becomes large (see Eq.~\ref{eq:pfilm}), with the result that the peak intensity near completion of odd-numbered layers is large compared to the starting intensity. Finally, Fig.\ref{fig:xray_schem}D simulates growth on SiO$_2$ as observed at the quarter-Bragg position ($L=0.25$).

A clear feature of the solid curves in the left-hand plots in  Fig.~\ref{fig:xray_schem} is that the scattering amplitude of a rough film approaches the center of the circle corresponding to ideal SF growth. For $L=0.5$, it has been shown previously \cite{Krause:2004ss} that the scattered intensity of a \textit{rough} film coincides precisely with that of a film with precisely 0.5 ML coverage, i.e. $\theta_1=0.5$, $\theta_{n>1}=0$. Figure~\ref{fig:xray_schem} shows that this coincidence does not occur for $L\neq0.5$. A more general interpretation of this limit is to view the scattering as arising from the different interfaces in the system\cite{Sinha:1988qb}. When the film/vacuum interface becomes rough, its effective scattering strength drops. Explicitly, evaluating Eq.~\ref{eq:xray6} for a rough film (e.g. taking $\theta_n = e^{-\delta n}$) leads to 
\begin{equation}
\label{eq:xray_rough}
A(q_z) \propto \left [ (\rho_0-\rho_1)e^{i 2\pi L\tau} + (\rho_1-\rho_2) \right ]\text{.}
\end{equation}
Thus, the scattering amplitude of a rough film, which corresponds to the center of the circular trajectories of ideal SF growth in the left-hand plots in Fig.~\ref{fig:xray_schem}, also corresponds to the scattering amplitude of the smooth, buried interfaces alone.  

\subsection{Layer-wise Rate Equation Model of Epitaxial Growth} \label{sect:models}

Even if the scattering parameters described in section \ref{sect:xray} are precisely known, it is clear that the coverages $\theta_n(t)$ at time $t$ are not uniquely determined by a single intensity value $I(t)$. To determine these coverages, additional information about the surface evolution is required. The most common approach is to construct a deterministic, parameterized model for the evolution of coverages $\theta_n(t)$. Such models take the form of coupled, differential equations:
\begin{equation}
\label{eq:rate1}
\frac{d\theta_n}{dt}=F(k_j,..., \theta_{n-1}, \theta_n, \theta_{n+1}, ..., t) \text{,}
\end{equation}
where the functions $F$ represent the additional information imposed on the system, and may or may not depend explicitly on a subset of coverages and time $t$. The parameters $k_j$ determine the $\theta_n(t)$'s, which are substituted into Eq. \ref{eq:xray2} to calculate $I(t)$. 

A large variety of models of the form in Eq.~\ref{eq:rate1} have been developed over the last several decades. Most of these originated and evolved in the context of specific experimental techniques, such as Auger Electron Spectroscopy (AES) \cite{Barthes:1981ij,Argile:1989sp, Mroz:2000oz,Fu:2003cc}, ion-beam assisted deposition\cite{Koponen:2000dk,Sillanpaa:2001ft,Huhtamaki:2007fp}, and RHEED \cite{Cohen:1989cp,Kariotis:1989gc}.  Models used to analyze XRR data \cite{Vandervegt:1992va,Woll:1999at,Mayer:2004hs,Kowarik:2009ie} as described here have virtually all drawn directly from those of Ref.~[\onlinecite{Cohen:1989cp}]. More recently, Trofimov \etal\cite{Trofimov:1997tw,Trofimov:1999yb,Trofimov:2003jw,Trofimov:2007ye,Trofimov:2008qf}, developed a new variant of this class of models, specifically for the purpose of linking atomistic kinetics to morphology in thin, multilayer films.

It is convenient to categorize these models according to how much they attempt to incorporate atom-level kinetics. Models by Kariotis \etal\cite{Kariotis:1989gc}, Trofimov \etal\cite{Trofimov:1997tw,Trofimov:1999yb,Trofimov:2003jw,Trofimov:2007ye,Trofimov:2008qf}, and Koponen \cite{Koponen:2000dk,Sillanpaa:2001ft} draw explicitly on nucleation theory \cite{Venables:1984}, and include additional equations to Eq.~\ref{eq:rate1} representing the adatom and island densities on each layer. The next simplest models, most notably those introduced by Cohen \etal\cite{Cohen:1989cp}, but including Refs.~[\onlinecite{Mroz:2000oz}], [\onlinecite{Fu:2003cc}], and [\onlinecite{Huhtamaki:2007fp}], do not explicitly attempt to model atom-level kinetics, but approximate these kinetics in the form of Eq. \ref{eq:rate1}. For example, the ``distributed'' model described in Ref.~[\onlinecite{Cohen:1989cp}] includes a mean-field representation of the step-density of a layer as a function of coverage $\theta$, which in turn controls the amount of downhill interlayer diffusion. Finally, Braun \etal\cite{Braun:2003ld} have developed a version of Eq.~\ref{eq:rate1} which attempts, explicitly, to avoid modeling atomistic kinetics. Rather, the equations Eq.~\ref{eq:rate1} are coupled via intermediate functions $J_n(t)$, chosen to be as simple as possible consistent with producing physically reasonable results for the functions $\theta_n(t)$.

Below, we describe implementations of three of the models described above: the empirical model described by Braun \etal\cite{Braun:2003ld}, a simplified version of the model introduced by Trofimov \etal\cite{Trofimov:1997tw}, and our variant \cite{Hong:2008gc,Amassian:2009mk,Amassian:2009yk} of the distributed model introduced in Ref.~[\onlinecite{Cohen:1989cp}]. In each case, we introduce analogous modifications that were found necessary to fit data obtained for our model systems, pentacene/SiO$_2$ and DIP/SiO$_2$. Specifically, successful fits, described in section~\ref{sect:results}, require that the kinetic parameters change with layer number for at least the first three layers. For the case of DIP/SiO$_2$, it is additionally required that the sticking coefficient of incident molecular species vary with coverage of the first layer.

First, we present two modifications to the so-called distributed model of Ref.~[\onlinecite{Cohen:1989cp}] that widen its range of applicability and connect, within the limits of the mean-field approach, to well-defined physical quantities. This modified Cohen (mC) model, (depicted schematically in Fig.~\ref{fig:real_schem}), is written:
  \begin{equation}
    \label{eq:distributed}
    \frac{d\theta_n}{dt}=R_n \left (1-\alpha_{n-1} \right ) \left ( \theta_{n-1} - \theta_n \right ) + R_{n+1} \alpha_n \left ( \theta_n - \theta_{n+1} \right ) \text{,}
  \end{equation}
where $R_n$ and $\alpha_n$ are the net deposition rate from vapor and downward interlayer transport probability into the n$\supth$ layer. The parameters $R_n$ and $\alpha_n$ implicitly represent all of the kinetic processes involving molecular attachment and transport, respectively. Rather than Cohen's original form for $\alpha_n$, we use 
\begin{equation}
\begin{split}
  \label{eq:interlayer}
  \alpha_n(\theta_n, \theta_{n+1}) & =\frac{k_-d(\theta_n)}{k_- d(\theta_n)+k_+ d(\theta_{n+1})} \\
& = \frac{e^{-E'}d(\theta_n)}{e^{-E'}d(\theta_n) + d(\theta_{n+1})} \mbox{,}
\end{split}
\end{equation}
where $d(\theta)$ represents the average step-edge density of a layer with coverage $\theta$. The factors $k_{-(+)}$ represent the rate of molecular attachment at downhill (uphill) steps, and $e^{-E'}=k_-/k_+$.  In the assumption of irreversible attachment, $E'=E_\text{ES}/kT$, where $E_\text{ES}$ is the Ehrlich-Schwoebel barrier for interlayer transport\cite{Bromann:1995fd}. In general, the function $d(\theta)$ could depend on layer number, for example if the island density changes as a function of layer thickness or island shape. Since such differences are virtually indistinguishable from differences in $E'$, we use the same function $d(\theta)$ for each layer, but typically allow variations in $E'$ as a function of $n$.  

The rate equations described by Eq.~\ref{eq:rate1} do not explicitly prevent overhangs, i.e. solutions in which $\theta_n(t) > \theta_{n-1}(t)$ for some $n$ and $t$. Overhangs can occur, for example, if $\alpha_n>\alpha_{n-1}$, so that the interlayer transport into layer $n$ is much greater than the transport of layer $n$ into layer $n-1$. In our implementation of the Cohen model, overhangs are prevented by forcing $\alpha_n$ to approach 0 as $\theta_n$ approaches $\theta_{n-1}$.

The parameterization in Eq.~\ref{eq:interlayer} fixes a drawback of the original form, in which $\alpha_n$ is defined as\cite{Cohen:1989cp} $\alpha_n = A d(\theta_n)/(d(\theta_n)+d(\theta_{n+1}))$. In this case, $A=1$ corresponds to $E' = -\infty$ in our model, whereas $0\leq A<1$ corresponds, approximately, to $E'>0$. The region that is mathematically forbidden in Cohen's original form, $-\infty<E'\leq 0$, is precisely the region of parameter space where extended LBL oscillations are expected. As noted in Ref.~\onlinecite{Cohen:1989cp}, that model is unable to reproduce such behavior. 

Our second modification to the Cohen model is to employ recent work of Tomellini \etal\cite{Tomellini:2006va} concerning the form of $d(\theta)$: the mean-field step-edge density of a growing film comprised of 2D islands. Two forms of $d(\theta)$ are obtained, depending on whether or not islands rearrange when they merge, reducing their total perimeter. The zero and complete rearrangement limits are labeled ``impingement'' and ``coalescence'' regimes, respectively, and are found to have step densities $d_{\text{im}}$ and $d_{\text{co}}$\cite{Tomellini:2006va}:
\begin{equation}
 \label{eq:tomellini_dendritic}
 d_{\text{im}}(\theta)=2\sqrt{\pi N_0} (1-\theta) \left [ \ln\frac{1}{(1-\theta)} \right ]^{1/2}\text{,}
\end{equation}
\begin{equation}
        \label{eq:tomellini_compact}
        d_{\text{co}}(\theta)=\sqrt{\theta (1-\theta)}\exp\left ( - \frac{\theta}{2(1-\theta)}\right ) \text{.}
 \end{equation}
Equation \ref{eq:tomellini_compact} reflects the fact that if islands rearrange upon coalescence, the density of holes remaining in the surface near layer completion is smaller than the island density at nucleation. This implies that the step-edge density is also smaller, e.g. $d(1-\epsilon) \ll d(\epsilon)$. Real systems are expected to exhibit behavior intermediate between that of Eqs.~\ref{eq:tomellini_dendritic} and \ref{eq:tomellini_compact}. In practice, we parameterize this form so that the degree of coalescence may be tuned by a parameter $\parcfrac$, $0 \le \parcfrac \le 1$:
\begin{equation}
\label{eq:d_fit}
d(\theta)=\parcfrac d_{\text{co}}(\theta) + (1-\parcfrac) d_{\text{im}}(\theta) \text{.}
\end{equation}

In order to solve Eq.~\ref{eq:distributed} using Eqs.~\ref{eq:interlayer} and \ref{eq:d_fit}, it is necessary to ``nucleate''  layers 2 and above, since $\alpha_n$ otherwise remains equal to one, precluding nucleation of layer $n+1$.  This is done by computing $d(\theta_{n+1})$ using a small value ($\epsilon \sim 10^{-5}$) of $\theta_{n+1}$ when $\theta_n$ exceeds a critical coverage $\theta_{n,\text{cr}}$. Explicitly:
\begin{equation}
 \alpha_n = \begin{cases} 1 \text{  ,  } \theta_n<\theta_{n,\text{cr}} \\
\alpha_n (\theta_n, \theta_{n+1} + \epsilon) \text{  ,  } \theta_n \ge \theta_{n,\text{cr}} \text{.}\\
\end{cases} 
\end{equation}

Using Eqs.~\ref{eq:interlayer} and~\ref{eq:d_fit}, results obtained by numerically solving Eq.~\ref{eq:distributed} may be tuned, via the parameter $E'$, from 3D growth ($E' \gtrsim 1$) to perfect LBL growth ($E' \lesssim -1$). More complicated behavior can result from allowing different values of $E'$ for different layers. For pentacene deposition on SiO$_2$, we find that reasonable results (discussed further in section~\ref{sect:results}) are obtained by employing several values of $E'$, e.g. $E_1$, $E_2$ for the first two layers, followed by asymptotic approach to $E_2+\Delta E_N$ for layers $n\ge3$, e.g. $E_{n\ge3} = E_2 + \Delta E_N\times\exp((n-2)/N_0)$. For DIP/SiO$_2$ and other systems exhibiting growth rate acceleration, $R_1$ is allowed to differ from $R_{n>1}$, simulating a difference in sticking coefficient for molecules incident on the growing film as opposed to the bare substrate. In summary, then, our model for real systems involves ten growth parameters, $R_1$, $R_{n>1}$, $E_1$, $E_2$, $\Delta E_N$, $\theta_{1,\text{cr}}$, $N_0$, $\theta_{2,\text{cr}}$, $\theta_{N,\text{cr}}$, and $\delta$. 

The second model in our comparison, introduced by Ref.~[\onlinecite{Braun:2003ld}] and referred to below as the Braun/Kaganer (BK) model, is explicitly simpler than those described in Ref.~[\onlinecite{Cohen:1989cp}]. Equation \ref{eq:rate1} takes the form:
\begin{equation}
\label{eq:braun1}
\frac{d\theta_n}{dt}=R_n(J_n-J_{n+1})\text{,}
\end{equation}
where the intermediate functions $J_n$ depend explicitly on time as 
\begin{equation}
\label{eq:braun2}
J_n=\frac{1}{2} \left ( 1+\tanh \left (\frac{t-t_n}{\beta_n} \right )  \right )   \text{.}
\end{equation}
Here, the times $t_n$ are determined directly by the rates $R_n$, while the parameters $\beta_n$, which are generally chosen to increase monotonically\cite{Braun:2003ld}, have the effect of determining the film roughness.  As with our model above, we find that additional variation in the values of $\beta_n$ for the first few layers is required to obtain a good fit to data. We define $\beta_1$ and $\beta_2$ as free parameters, while for layers $n\ge3$, $\beta = \beta_3(n-2)^\alpha$. Because of the explicit time dependence of the intermediate functions $J_n$, this model does not strictly allow for a coverage dependent growth rate. Nevertheless, allowing different values for $R_n$ does approximate such behavior. As with the mC model, the value of $R_1$ is allowed to differ from that of $R_{n>1}$ for systems exhibiting growth rate acceleration. To calculate the $t_n$'s, we make the explicit assumption that layer $N$ is completed at time $t_N = 1/R_1 + \sum_{n=2}^N1/R_{n>1}$, leading to $t_n = n R_{n>1}/(R_1 R_{n>1})$. Thus, the BK model includes six parameters, $R_1$, $R_{n>1}$, $\beta_1$, $\beta_2$, $\beta_3$, and $\alpha$. 

The last model employed for our comparison is a simplified version of an atomistic, rate-equation model developed by Trofimov \etal~\cite{Trofimov:1997tw,Trofimov:1999yb,Trofimov:2003jw,Trofimov:2007ye,Trofimov:2008qf}. Unlike the layer-wise models just described, this model includes three rate equations for each layer, describing the rates of change of the adatom density, island density, and coverage. Downhill transport is controlled by a so-called ``feeding zone'' (FZ) $\xi_n$ of each layer $n$: atoms that are incident onto the feeding zone $\xi_n$ of layer $n$ remain on that layer and hence contribute to layer ${n+1}$.  Adatoms that land on top of layer $n$ but outside this feeding zone diffuse downward, thus increasing $\theta_n$. A clear advantage of this model is the ability to directly connect morphology, including in-plane parameters such as the nucleation density, evolves as a function of layer number, and as a function of real, physical parameters such as $D/J$, the ratio of adatom diffusivity to the incident flux. For example, it is found that, even for homoepitaxy, the saturation island density decreases as a function of layer number during LBL growth\cite{Trofimov:2003jw}. This phenomena has recently been directly observed in SrTiO$_3$ homoepitaxy using pulsed laser deposition \cite{Ferguson:2009ig}. 

For the limited problem of examining layer coverages, the FZ model may be re-parameterized so that it is represented by just one equation per layer. Examination of Ref.~[\onlinecite{Trofimov:2003jw}] shows that all of the atomistic physics is contained in the evaluation of the critical coverage for each layer $\theta_{n,\text{cr}}$. Defining this as a fit parameter, we have:
\begin{equation}
    \label{eq:trofimov}
    \frac{d\theta_n}{dt}=\begin{cases} R_1 (1-\theta_1) + R_{n>1} (\theta_1-\xi_1) \text{  ,  } n=1 \\
 R_{n>1} (\xi_{n-1}-\xi_n) \text{  ,  } n > 1 \text{,} 
\end{cases}
\end{equation}
where
\begin{equation}
  \label{eq:feedzone}
\xi_n = \begin{cases}0 \text{  ,  } \theta_n < \theta_{n,\text{cr}} \\
1 - e^{ -\left [ \sqrt{-\ln (1-\theta_n)} - \sqrt{-\ln(1-\theta_{n,\text{cr}})} \right ]^2} \text{  ,  } \theta_n \ge \theta_{n,\text{cr}} 
\text{.}
\end{cases}
\end{equation}
As in the models above, we have incorporated variation in the sticking coefficient of incident species through the parameters $R_1$ and $R_{n>1}$, and obtain reasonable fits by allowing $\theta_{n,\text{cr}}$ to change with layer number. The analysis shown below incorporates six growth parameters, $R_1$, $R_{n>1}$, $\theta_{1,\text{cr}}$, $\theta_{2,\text{cr}}$, $\theta_{\infty,\text{cr}}$, and a parameter $N_0$ determining the asymptotic approach to $\theta_{\infty,\text{cr}}$.

\section{Comparison to Experiment}
\label{sect:results}

\subsection{Pentacene/SiO$_2$} \label{sect:pent}

The AFM results in Fig.~\ref{fig:pent_raw}B-E obtained by hyperthermal growth are consistent with those of prior, voluminous work on (thermal) pentacene/SiO$_2$ MBE\cite{Ruiz:2003um,Ruiz:2004su,Mayer:2004hs, Mayer:2006da,Hong:2008gc}. Growth begins in a LBL mode, so that the first layer is nearly complete before the second layer nucleates. Subsequently, roughness increases quickly, resulting in a late-time morphology characterized by large, multilayer islands. The precise nature of this transition varies with growth conditions\----particularly temperature \cite{Mayer:2006da}. 

\begin{figure}
\includegraphics[scale=1]{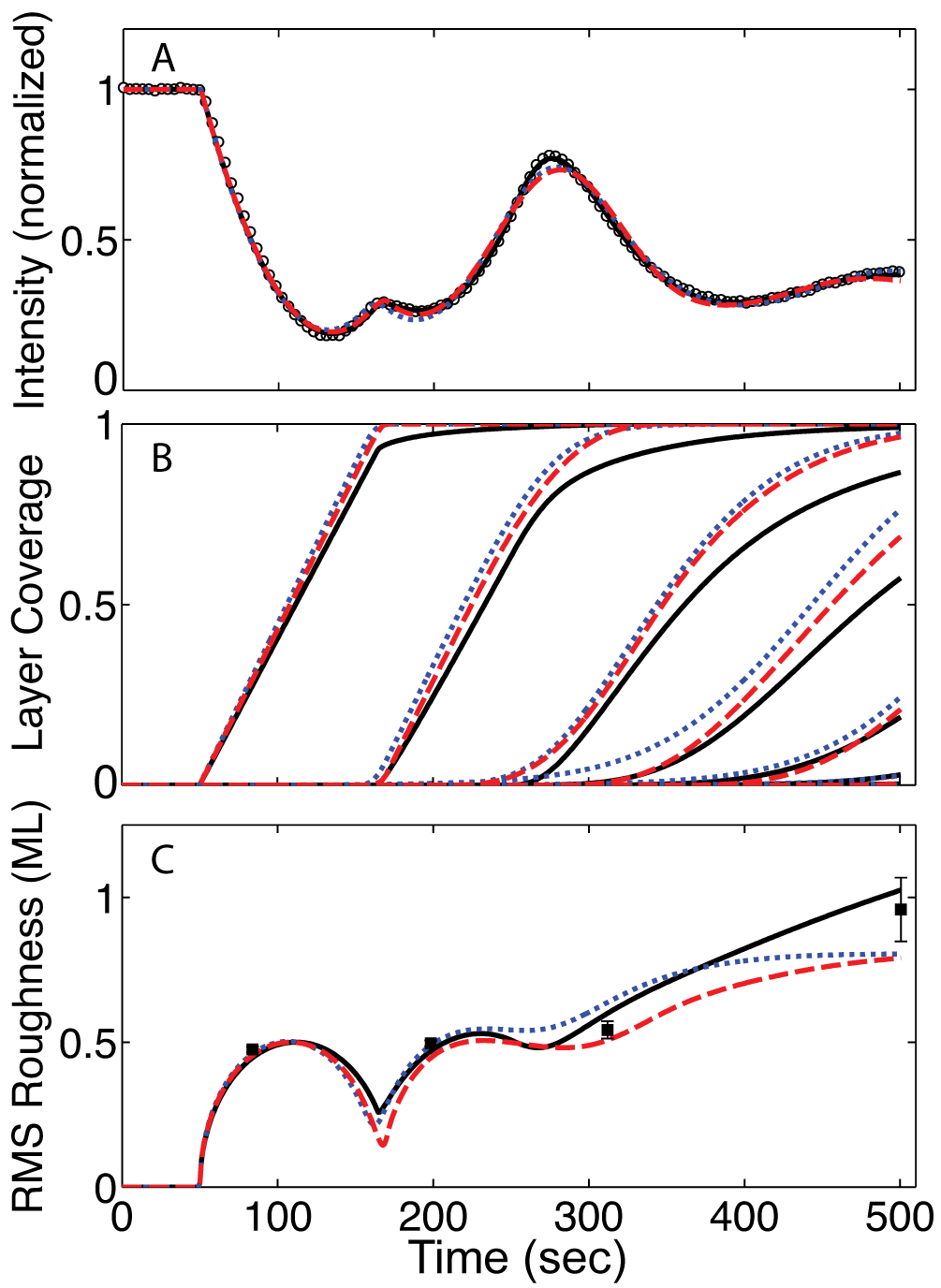}
\caption{\label{fig:pent_xray} (Color online) a) (open circles) XRR data obtained near the pentacene anti-Bragg position during pentacene deposition on SiO$_2$. Also shown are best-fit curves to the mC (solid line), BK (dotted line), and FZ (dashed line) models described in the text. b) Layer coverages from the three models shown in (a). c) Evolution of root-mean square roughness obtained from the three models shown in (a) and (b). Filled squares are measured rms roughness values obtained from the AFM images shown in Fig.~\ref{fig:pent_raw}B-E, corresponding to film thicknsesses of 0.34, 1.4, 2.5, and 4.4 ML. }
\end{figure}

Figure \ref{fig:pent_xray}A reproduces the data from the thickest film shown in Fig.~\ref{fig:pent_raw}, along with fits to the three models described in section \ref{sect:models}. Figures~\ref{fig:pent_xray}B-C show the evolution of layer coverages and rms roughness, respectively, resulting from the fits in Fig.~\ref{fig:pent_xray}A, in addition to the rms roughness values (Fig.~\ref{fig:pent_xray}C) obtained from AFM on each of the four films. Error bars in Fig.\ref{fig:pent_xray}C, visible only for the thickest film, correspond to standard deviations of values obtained from different quadrants of the images in Fig.~\ref{fig:pent_raw}B-E. As discussed further below, however, as much as 15\% variation in growth rate was subsequently found across the 1-cm wide substrate, all of which is sampled by the x-ray beam. Therefore, these error bars may underestimate the variation in film morphology as observed by the x-ray measurements.
Nevertheless, reasonably good agreement is also obtained in rms roughness evolution among the models and the AFM results, particularly at early times. 

The parameters for the fits in Fig.\ref{fig:pent_xray} are shown in Table~\ref{tab:pent_pars}. Since deposition was performed directly on clean SiO$_2$, as confirmed by x-ray reflectivity prior to growth, no interfacial layer was included ($\rho_1=0$, $\tau=0$). For $\rho_0$, a bulk mass density of 2.2 g/cm$^3$ was used, corresponding to the assumption that electron density is proportional to mass density. For this case, no growth rate acceleration is observed: allowing $R_{n>1}$ to vary from $R_1$ does not improve the fit. Thus, $R_{n>1} =R_1$ for the fits shown. All models reproduce the conspicuous features of the data, giving R-squared values of 99.4\% or higher. However, the fits are not perfect: the BKand FZ models give comparable quality fits, with $\chi_\nu^2$ values\footnote{In this paper, $\chi_\nu^2$ is the weighted sum of squared residuals, divided by the degrees of freedom.} of 35 and 34, respectively, indicating that the differences between the model and data are statistically significant\cite{num_rec_fort}. For the mC model, the best fit, with  $\chi_\nu^2=7$,  was obtained with $\Delta E_2$ and $\theta_{\infty,\text{cr}}$ at 0, and the remaining parameters as shown in Table~\ref{tab:pent_pars}. Error bars in the table correspond to an increase in $\chi_\nu^2$ by 1, after re-optimization of the remaining parameters.

\begin{table*}  
\begin{center}
\begin{tabular}{|c|c|c|c|c|c|}
\hline
\multicolumn{2}{|c|}{mC} & \multicolumn{2}{|c|}{BK} &\multicolumn{2}{|c|}{FZ} \\
\hline
$\rho_2$ & 1.458$\pm $0.03 & $\rho_2$ & 1.354$\pm$0.03 & $\rho_2$ & 1.373$\pm$0.02 \\
$\Delta$ & 0.202$\pm $0.004 & $\Delta$ & 0.172$\pm$0.004 & $\Delta$ & 0.1913$\pm$0.004 \\
$L$           & 0.4579$\pm$0.003 & $L$ & 0.482 $\pm$0.004 & $L$ & 0.467$\pm$0.004 \\
$R_1$ & 0.0081$\pm 2e^{-4}$ & $R_1$ & 0.0089$\pm 6e^{-5}$ & $R_1$ & 0.0086$\pm 5e^{-5}$ \\
 $\delta$ & 0.88 $\pm$ 0.07    & \multirow{3}{*}{$\beta_1$} & \multirow{3}{*}{ 6 $\pm$ 8} & \multirow{3}{*}{$\theta_{1,\text{cr}}$} &\multirow{3}{*}{ 0.88$\pm$0.2 }\\
$E_1$ & -0.55$\pm$0.2   & & & & \\
$\theta_{1,\text{cr}}$ & 0.93 $\pm$ 0.1 & & & & \\
$\theta_{2,\text{cr}}$ & 0.65$\pm$0.04 & $\beta_2$ & 38.8$\pm$3 & $\theta_{2,\text{cr}}$ & 0.341$\pm$0.02 \\
$\Delta E_N$ & 0.99$\pm$0.2 & $\beta_3$ & 87.0$\pm$3 & $\theta_{\infty,\text{cr}}$ & 0.078$\pm$0.005 \\
$N_0$ & 0.01+0.4/-0.01 & $\alpha$ & 0.0$\pm$0.1 & $N_0$ & 0.07 \\
\hline
\end{tabular}
\end{center}
\caption{\label{tab:pent_pars} Fit parameters for the three fits shown in Fig.~\ref{fig:pent_xray}} 
\end{table*}

The fact that the $\chi_\nu^2$ values differ significantly from 1 reflects, in part, the high accuracy of the data: the average intensity $\bar{I}$ in Fig.~\ref{fig:pent_xray}A is 2.4$\times$10$^4$, so that the mean statistical uncertainty in the data  $\left <\sqrt{I}/I\right >$ is 0.7\%. Alternatively, we compare the mean absolute residual with the mean of the data $\left < \left | I - I_m \right | / I_m \right >$. For the fits in Fig.~\ref{fig:pent_xray}, the mC model differs from the data by 1.4\%, the BK and FZ models by 3.2\%. 

\begin{figure}
\includegraphics[scale=1]{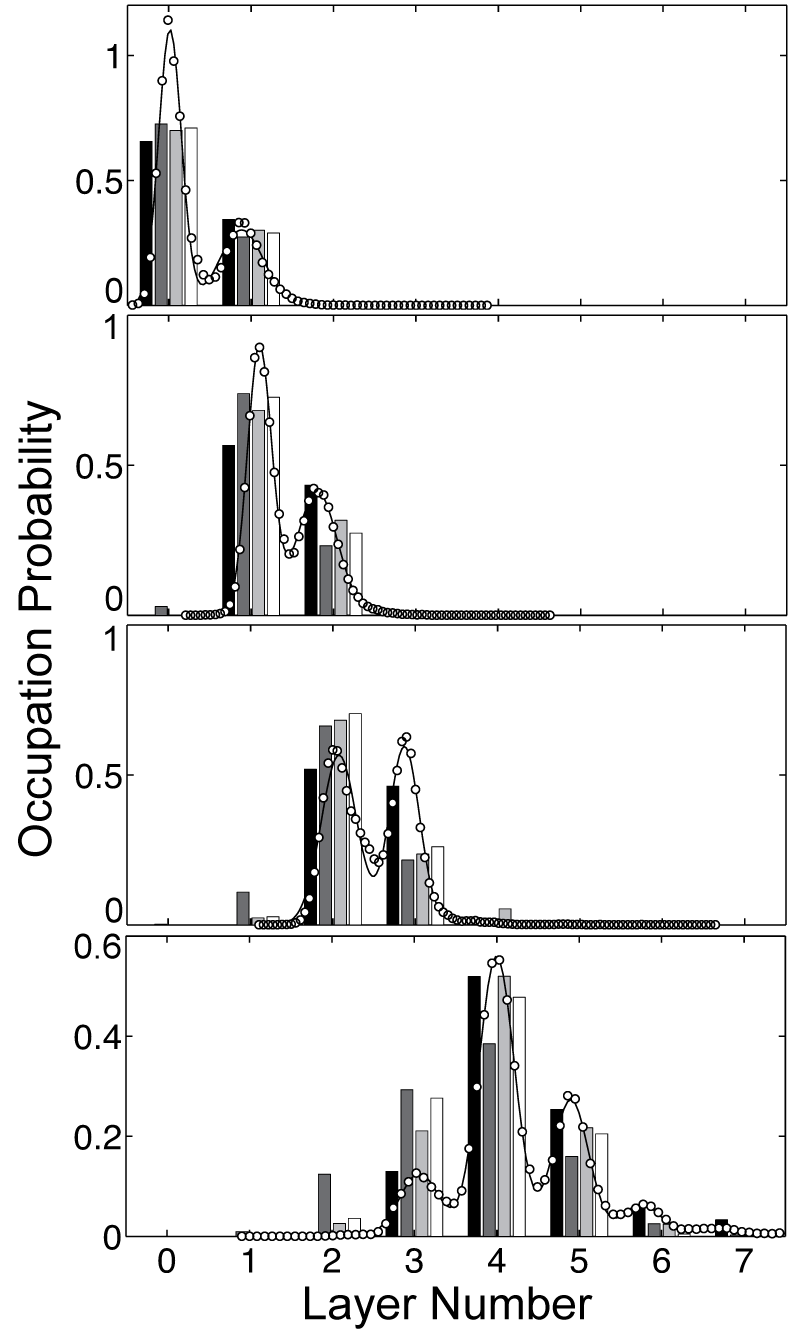}
\caption{\label{fig:pent_hists} Height distributions obtained from the AFM images in Fig.~\ref{fig:pent_raw} and from best-fit simulations shown in Fig.~\ref{fig:pent_xray}A. Open circles represent height distributions obtained directly from the AFM data, solid lines indicate fits of these data with a sum of several Gaussian distributions (one per layer), and black bars indicate the area of each Gaussian distribution so obtained. Dark grey, light gray, and white bars represent height distributions obtained from the mC, BK, and FZ models, respectively. }
\end{figure}

In view of the greater number of parameters in the mC model, the better fit is not unexpected. The question remains whether this statistically improved fit corresponds to more accurate layer coverages in Fig.~\ref{fig:pent_xray}B. The rms roughnesses appear to suggest that this is the case, since the measured rms roughness at $t=500$s corresponds more closely to the mC model than to the others. A more detailed comparison of the models and films is given in Fig.~\ref{fig:pent_hists}, which shows height distributions obtained by the AFM data in Fig.~\ref{fig:pent_raw} as well as the height distributions obtained from each of the best-fit simulations in Fig.~\ref{fig:pent_xray}. In each plot, discrete height distributions (represented as black bars) were obtained by fitting the continuous distribution to a sum of several Gaussian peaks, one per layer, and equating the area of each peak to fractional, exposed occupancies for each layer $c_n$.  If the $c_n$'s are normalized such that $\sum_n c_n = 1$, they are related to the layer coverages $\theta_n$ as: $c_n = \theta_n - \theta_{n+1}$. For AFM images, the indexing $n$ of layers is chosen such that the total thickness $\Theta = \sum_n n c_n = \sum_n \theta_n$ most closely matches thickness estimates provided either by the x-ray data or, alternatively, by the sub-monolayer growth rate. Although not shown, we assume a maximum error of 15\%  in the $c_n$'s, arising from the growth rate inhomogeneity across the film surface, as discussed above.  Figure~\ref{fig:pent_hists} reveals that the agreement between the mC model (dark gray) and AFM is not conspicuously better, and may in fact be worse than that of the other simulations. Thus, the lower value of  $\chi_\nu^2$  does not necessarily imply a more accurate representation of the true morphological evolution of the film. 

The parameters in Table \ref{tab:pent_pars} warrant several comments. First, the values of $\rho_2$ for all three fits are close to, but somewhat larger than that of bulk pentacene, 1.3 g/cm$^3$. The sign of this discrepancy is consistent with the expected error due to the uniform slab approximation. The fact that the electron density in pentacene is slightly concentrated around the molecular center (along the c axis) has the effect, at $L$ values below the first Bragg Peak, of increasing the scattering amplitude of this layer compared to a uniform slab with equivalent average density. A larger, \textit{uniform} electron density compensates for this difference. Second, although the data are nominally obtained at $L=0.5$ (the anti-Bragg position), good fits require $L\approx0.46-0.48$, and an offset between the substrate and the first layer of $\Delta_z = \Delta \times c \approx3$\AA. Possible contributions to the deviation of $L$ from 0.5 include the experimental uncertainty, defined by the detector slits to be $\Delta L=0.03$, and a change, with layer thickness, of the pentacene d-spacing $c$. Such a change has been reported by Fritz \etal~\cite{Fritz:2004id}. The offset $\Delta_z$ is assumed to correspond to a thin interfacial region at the SiO$_2$/film interface. The remaining parameters in Table \ref{tab:pent_pars}, those describing the growth morphology, all show effectively the same, monotonic trend from smooth to rough growth. Specifically, rougher growth corresponds to increasing values of $\beta_n$ in the BK model, decreasing values of $\theta_{n,\text{cr}}$ in the FZ model, and a combination of increasing $E_n$ and decreasing $\theta_{n,\text{cr}}$ in the mC model.

Finally, we note that the three models predict growth rates $R_1$ which vary from 0.0081 to 0.0089 ML/second. Evidently, even for the fairly straightforward problem of determining the growth rate of this simple system, the details of the model strongly affect the results. The origin of this variation is the link between the onset of roughness and the appearance of a peak. In general, extrema in the scattering data occur when the growth rate of one layer overtakes that of the layer below it. For perfect LBL growth, this moment coincides with layer completion, producing cusp-like peaks. But for a film undergoing roughening, this transition can occur at lower coverage for each consecutive layer, resulting in a shorter time between peaks than the time to deposite 1 ML. If the roughening transition is abrupt,  the growth rate can appear to accelerate. This is well illustrated by  Fig.~\ref{fig:pent_xray}A.  Growth begins at  $t=50$s, and the first two local maxima occur at $t=167$s and $t=275$s, corresponding to growth rate estimates of  0.0085 and 0.0093 ML/second. The three different models represented in Fig.~\ref{fig:pent_xray}A model the onset of roughness in slightly different ways, resulting in three different estimates of growth rate.

The variation in best-fit growth rates among different models, all of which fit the data well, raises the question of whether there is an independent means of measuring growth rate and, especially, of determining whether or not growth rate acceleration has occurred. We reiterate that for the hyperthermal growth method used here \cite{Schroeder:2004phd,Amassian:2009mk}, the use of a quartz crystal monitor is not possible due both to the narrowly-directed beam profile and the possibility that sticking coefficients are generally substrate-dependent. We present two alternatives. First, the cusp-like nature of the first local maximum in Fig.~\ref{fig:pent_xray}A suggests that this peak closely coincides with completion of the first layer. Thus, \textit{provided} that we are confident that growth rate acceleration does not occur, and that the surface morphology does not develop large height asymmetries during the first layer\cite{Headrick:1998kw}, the growth rate estimate given by the peak, 0.0085 ML/second should provide an accurate estimate of the growth rate. This growth rate falls within the results from the three fits. 

\begin{figure}
\includegraphics[scale=0.5]{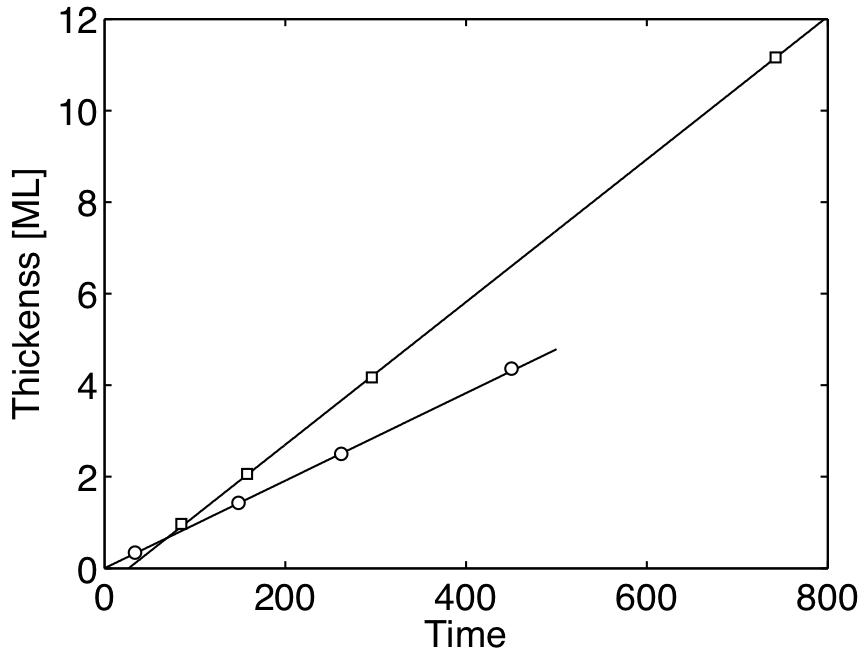}
\caption{\label{fig:afm_thick} Thickness, determined by AFM as described in the text, vs. time of (circles) pentacene films represented in Fig. \ref{fig:pent_raw} and (squares) DIP films represented in Fig.~\ref{fig:dip_afm}. Also shown are linear fits to the AFM-determined thicknesses. The best-fit line to the pentacene data intersects the origin, indicating an absence of growth rate acceleration, and a growth rate of 0.0096 $\pm$ 0.0002 ML/s. For DIP, only the later three point are used for the fit, yielding a late-time growth rate of $R_{n>1}$ of 0.0156 $\pm$ 0.001 ML/s, and a negative intercept, indicating growth rate acceleration.}
\end{figure}

A second alternative measure of the growth rate can be made using the height distributions represented in Fig.~\ref{fig:pent_hists} to obtain a plot of thickness vs. time. As noted above, if layers in an AFM image are correctly indexed, the thickness $\Theta$ is equivalent to the center-of-mass of the height distributions $\sum_n n c_n$. The circles in Figure \ref{fig:afm_thick} represent the thicknesses of the four films in Fig.~\ref{fig:pent_raw} obtained in this manner, along with the best-fit line to those data. Clearly, a line describes the data well. Unfortunately, the growth rate obtained in this fashion, 0.0096$\pm$0.0002 ML/s, does not fall within the growth rates obtained from fits, and is not consistent with 0.0085 ML/s obtained by simple inspection of Fig.~\ref{fig:pent_xray}A. It is difficult to reconcile the x-ray data with a growth rate of 0.0096 ML/s, since this growth rate would imply that the sharp, cusp-like peak at $t=167$s occurs at a thickness \textit{greater } than 1 ML. 

One possible source of error of the AFM-obtained thicknesses in Fig.~\ref{fig:afm_thick} is the finite size of the AFM tip. The finite spatial resolution resulting from tip size will always result in an \textit{overestimate} of the film thickness, since holes will appear smaller than their true size, while islands will appear larger than their true size. We estimated this error by finding the total perimeter of all of the islands for several layers and images in Fig.~\ref{fig:pent_raw}B-E, and multiplying by the estimated error, ~4 nm, of the measured, lateral position of the step due to a finite tip radius of 10 nm. This lateral position is assumed to correspond to half of the full 1.5nm height of a single ML step, and the AFM tip radius is obtained from SEM measurements of a nominally identical tip to that used to obtain the images in  Fig.~\ref{fig:pent_raw}B-E. The resulting error is approximately 1\% of the total image area, much less than the 10\% discrepancy between the simulations and Fig.~\ref{fig:afm_thick}. Of course, because the islands are fractal,  our measurement of island perimeter is an underestimate. A simple estimate of this error, considering the AFM tip size, molecule size, and estimated fractal dimension $d_f\approx1.8$,\cite{Ruiz:2003nw} suggests our perimeter measurement to be in error by no more than a factor of 2.  

Another explanation for this discrepancy is that the growth rate is not uniform over the portion of the film measured by x-rays. Because these measurements are performed at an incidence angle of ~1$^\circ$, the 0.5mm x-ray beam probes the entire 10 mm width of the substrate. Subsequent AFM measurements (not shown) confirm that the growth rate near the edge of the film is 15\% smaller than that near the center. Thus, the average growth rate measured by the x-ray beam is smaller than that at the center of the film, where the measurements in Fig.~\ref{fig:pent_raw} were obtained.

Regardless of the disagreement in growth rate implied by the x-ray and AFM analysis, the two approaches agree with regard to the absence of growth rate acceleration. Referring again to Fig.~\ref{fig:afm_thick}, this conclusion comes from the fact that the best-fit line of the pentacene data intersects the origin to within 0.02 ML. This finding contrasts the case of DIP/SiO$_2$, also represented in Fig.~\ref{fig:afm_thick} and discussed in detail below. 

\subsection{DIP/SiO$_2$} \label{sect:dip}

\begin{figure}
\includegraphics[scale=0.5]{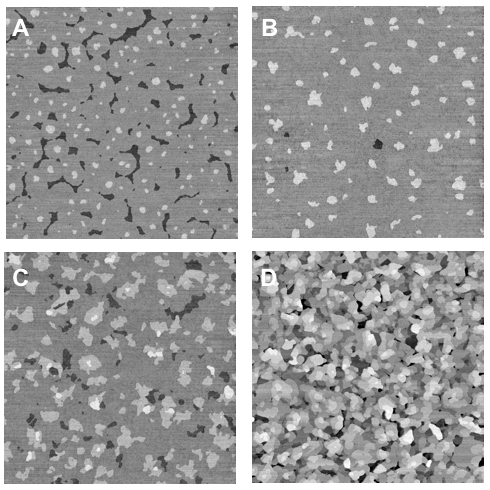}
\caption{\label{fig:dip_afm} AFM images of four thin films of DIP/SiO$_2$ of increasing thickness grown in immediate succession on four areas of the same substrate: (A) 0.97 ML; (B) 2.06 ML; (C) 4.17 ML; (D) 11.16 ML. All images are 10$\mu$m$\times$10$\mu$m}
\end{figure}

\begin{figure}
\includegraphics[scale=1]{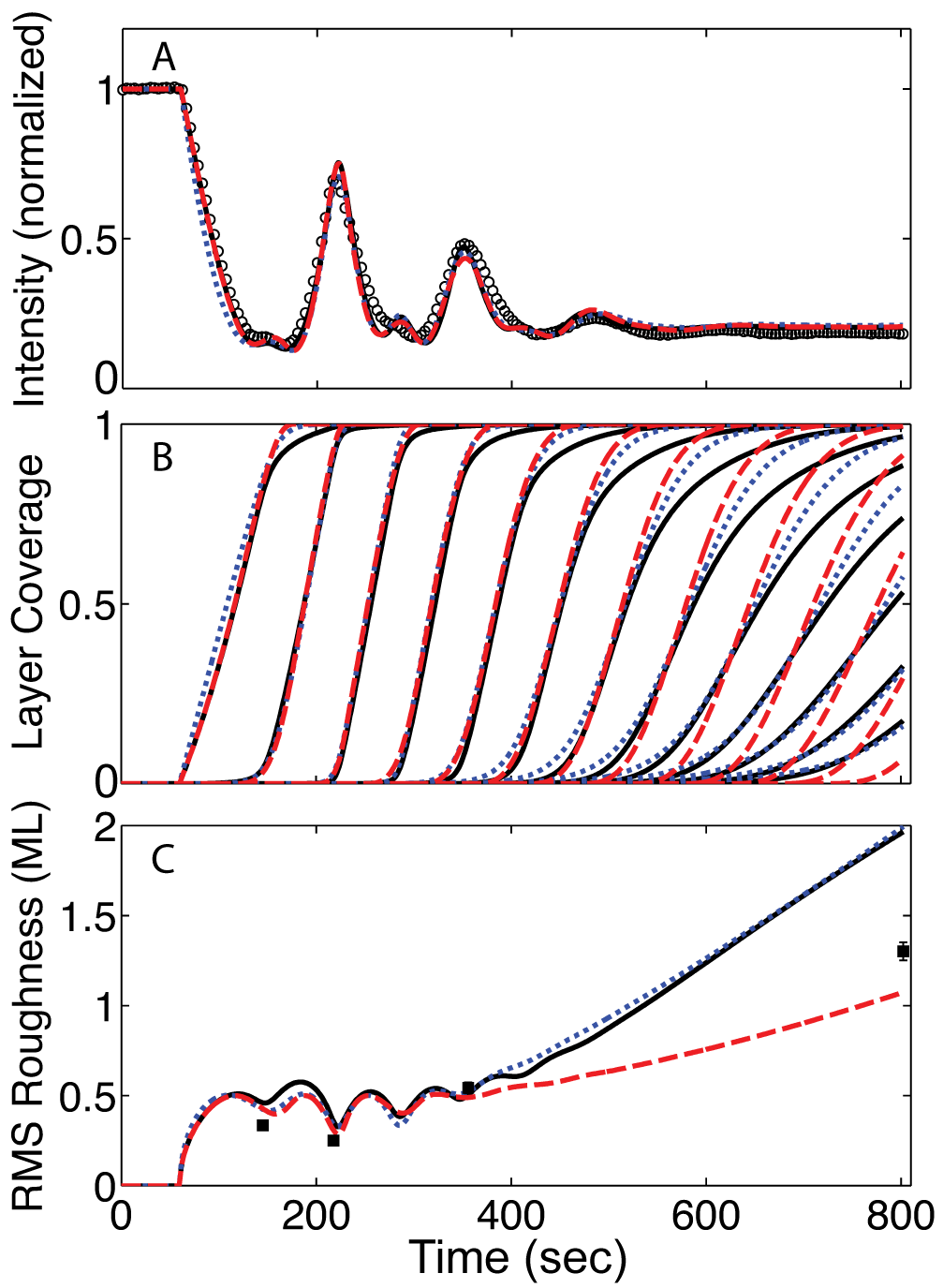}
\caption{\label{fig:dip_xray} (Color online) As in Fig.~\ref{fig:pent_xray} , a): (open circles) XRR anti-Bragg data during DIP deposition on SiO$_2$, along with best-fit curves from the mC (solid line) , BK (dotted line), and FZ (dashed line) models. b) Layer coverages from the three models shown in (a). c) Evolution of root-mean square roughness obtained from the three models shown in (a) and (b), with values obtained from ex situ AFM shown as  filled squares.}
\end{figure}

Figure \ref{fig:dip_afm} shows AFM data obtained from a thickness series of DIP/SiO$_2$. As in Fig.~\ref{fig:pent_raw}, the films were grown in immediate succession on the same substrate, but at a substrate temperature of 89$^\circ$C. This increased temperature may contribute to the two most conspicuous differences between Figs.~\ref{fig:dip_afm} and \ref{fig:pent_raw}, namely, that DIP exhibits compact, rather than dendritic island morphology, and more persistent LBL growth. The later observation is clearly seen by comparing Fig.~\ref{fig:pent_xray}A to Fig.~\ref{fig:dip_xray}A, which shows \insitu x-ray data obtained for the film shown in Fig.~\ref{fig:dip_afm}D.

The squares in Fig.~\ref{fig:afm_thick} represent thickness vs. time of the four films in Fig.~\ref{fig:dip_afm}, obtained by combining data in Fig.~\ref{fig:dip_afm} with qualitative analysis of \insitu x-ray data acquired during growth. In contrast to the pentacene case, a best-fit line to these four points clearly has a negative intercept, indicating growth rate acceleration. Because of this, only the later three points are used for the fit in Fig.~\ref{fig:afm_thick}, yielding a late-time growth rate of 0.0156 $\pm$ 0.001 ML/s. 

To determine of the degree of growth rate acceleration, defined as $(R_{n>1}-R_1)/R_1$, we generally rely on quantitative analysis of XRR data. However, for the special case of LBL growth, this, like the growth rate, can be obtained directly from the data in Fig.~\ref{fig:afm_thick}.  In this case, the rate of growth of the first layer is described by: 
\begin{equation}
\frac{d\theta_1}{dt} = R_1 (1 - \theta_1) + R_{n>1} \theta_1,
\end{equation}
and the remaining layers grow according to
\begin{equation}
\frac{d\theta_n}{dt} = R_{n>1} (1 - \theta_n).
\end{equation} 
Solving this system yields the total thickness $\Theta =\sum_n \theta_n$ as a function of time:
\begin{equation} \label{eq:accel}
\Theta(t) = \begin{cases}\frac{R_1}{R_{n>1}-R_1} \left ( e^{(R_{n>1}-R_1)t} - 1 \right )\text{  ,  } t\le t_1 \\
1+R_{n>1}(t-t_1) \text{  ,  } t > t_1\text{,} \\
\end{cases} 
\end{equation}
where $t_1=\log ( (R_{n>1}-R_1)/R_1 + 1)/(R_{n>1}-R_1)$. We now identify the linear regime in Eq.~\ref{eq:accel} with the best-fit line to the late-time thickness data in Fig.~\ref{fig:afm_thick}. Setting the intercept $b$ of that line equal to $1-R_{n>1}t_1$ yields 
\begin{equation}\label{eq:r1}
(1-b) \left ( \frac{R_1}{R_{n>1}} - 1 \right ) = \log \left ( \frac{R_1}{R_{n>1}} \right ),
\end{equation}
which can be numerically solved for $R_1$. For the DIP data in Fig.~\ref{fig:afm_thick},  $R_{n>1}=0.0156$ ML/s and $b=-0.417$ ML, yielding the result $R_1 = 0.0074$ ML/s. This corresponds to a growth rate acceleration of 111\%. 

In addition to the x-ray data in Fig.~\ref{fig:dip_xray}A,  Fig.~\ref{fig:dip_xray} shows fits of these data to the three models in section~\ref{sect:models}, the simulated layer coverages and rms roughnesses resulting from these fits, and rms roughnesses obtained from the AFM data in Fig.~\ref{fig:dip_afm}. In general, the modeled x-ray intensities in Fig.~\ref{fig:dip_xray}A show reasonable, but noticeably worse agreement than in Fig.~\ref{fig:pent_xray}A, and yield correspondingly worse $\chi_\nu^2$ values of 445, 756, and 509 for the mC, BK, and FZ models, respectively. As with the fits in Fig.~\ref{fig:pent_xray}A, we also compare the mean residual to the mean value of the data $\left < \left | I - I_m \right |/I   \right >$, obtaining values of 9.2\%, 12\%, and 10\%. The reasons for the difference in fit quality between Figs.~\ref{fig:pent_xray}A and~\ref{fig:dip_xray}A are unclear. However, previous work on DIP/SiO$_2$\cite{Kowarik:2009og} has demonstrated a clear structural change in a DIP thin film during growth of the first five layers. Such a change would have the effect of making some of the x-ray parameters discussed in section~\ref{sect:xray} time-dependent. 

\begin{figure}
\includegraphics[scale=1]{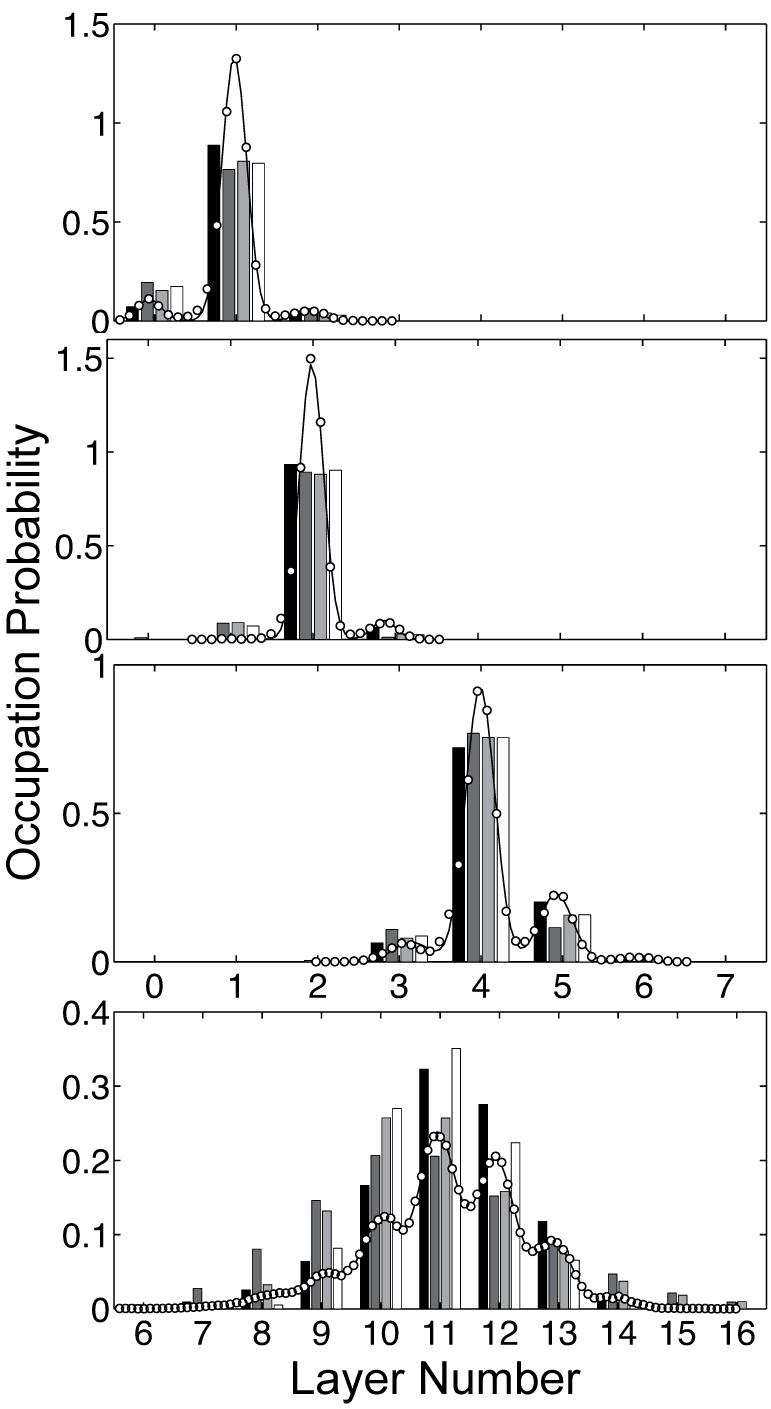}
\caption{\label{fig:dip_hists} Height distributions obtained from the AFM images in Fig.~\ref{fig:dip_afm} and from best-fit simulations shown in Fig.~\ref{fig:dip_xray}A. Open circles represent height distributions obtained directly from the AFM data, solid lines indicate fits of these data with a sum of several Gaussian distributions (one per layer), and black bars indicate the area of each Gaussian distribution so obtained. Dark grey, light gray, and white bars represent height distributions obtained from the mC, BK, and FZ models, respectively. }
\end{figure}

Referring to Fig.~\ref{fig:dip_xray}C, all three models compare very well with the AFM data up to $t=400$s, corresponding to a thickness of ~4 ML, beyond which the rms roughnesses predicted by the models diverge both from each other and from the actual film. At $t=800$s, corresponding to a thickness of 11.2 ML, the mC and BK overestimate the roughness, whereas the FZ model underestimates it. These observations are reflected in more detail in Fig.~\ref{fig:dip_hists}, which compares the actual height distributions obtained both from the AFM data and from the models. The correspondence between the simulated and measured height distributions for the three thinnest films is very good, whereas all three simulations depart significantly from the measured height distribution of the thickest film in the series.

\begin{figure}
\includegraphics[scale=1]{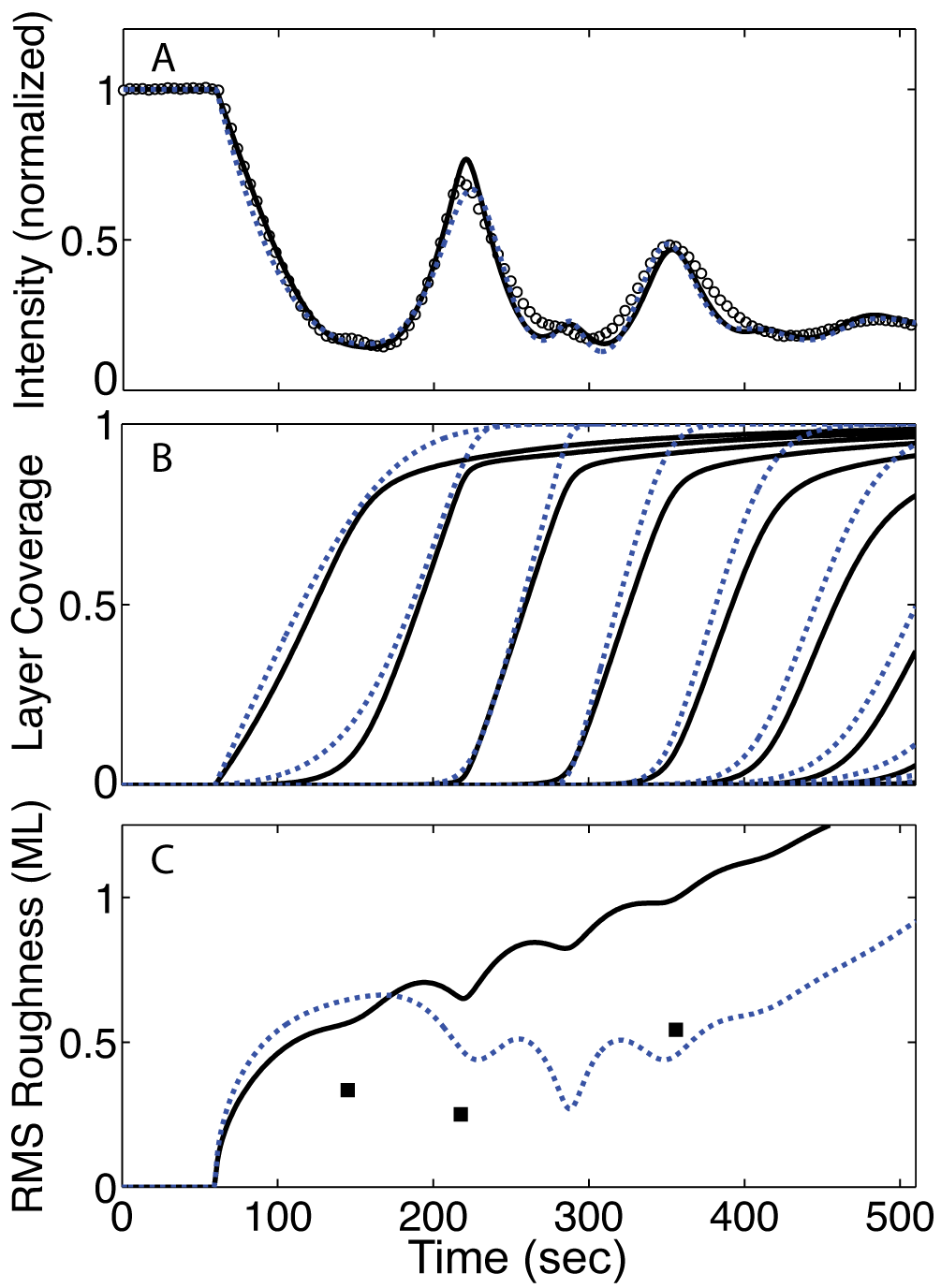}
\caption{\label{fig:dip_xray_alt} (Color online) As in Fig.~\ref{fig:pent_xray} , a): (open circles) The same XRR data as in Fig.~\ref{fig:dip_xray}, shown with alternative fits to the mC (solid line) and BK (dotted line) models.  b) Layer coverages from the two models shown in (a). c) Evolution of rms roughness obtained from the two models shown in (a) and (b), with values obtained from ex situ AFM (filled squares). These fits have lower $\chi_\nu^2$ values than those shown in Fig.~\ref{fig:dip_xray}, but rms roughnesses which are in clear disagreement with the AFM results.}
\end{figure}

Echoing the analysis of pentacene/SiO$_2$, the mC model, which yields the lowest value of  $\chi_\nu^2$ compared to the other two, nevertheless does not provide a conspicuously better representation of film growth. Moreover, both the mC and BK models tended towards second, deeper $\chi_\nu^2$ minima ($\chi_\nu^2 =$285, 417, respectively) characterized by the disappearance of the shallow local maximum near $t=150$s (see Fig.~\ref{fig:dip_xray_alt}A). As demonstrated by Fig.~\ref{fig:dip_xray_alt}C, the film morphology implied by these alternative fits are inconsistent with the AFM data. To avoid these minima, only a subset of parameters could be allowed to vary simultaneously, taking care that the best-fit model reproduced, at least weakly, all of the extrema in the data.

The parameters for the fits in Fig.~\ref{fig:dip_xray}A are shown in Table~\ref{tab:dip_pars}. As for the analysis of Fig.~\ref{fig:pent_xray}A, these fits presupposed the absence of an interfacial layer  ($\rho_1=0$, $\tau=0$), and employed $\rho_0=2.2$g/cm$^3$. For the mC model, $\theta_{1,\text{cr}}$, $\theta_{2,\text{cr}}$, and $\theta_{\infty,\text{cr}}$ were all fixed at 0. 

\begin{table*} 
\begin{center}
\begin{tabular}{|c|c|c|c|c|c|}
\hline
\multicolumn{2}{|c|}{mC} & \multicolumn{2}{|c|}{BK} &\multicolumn{2}{|c|}{FZ} \\
\hline
$\rho_2$ & 1.43$\pm $0.05 & $\rho_2$ & 1.40$\pm$0.06 & $\rho_2$ & 1.425$\pm$0.06 \\
$\Delta$ & -0.116$\pm $0.008 & $\Delta$ &-0.115$\pm$0.01 & $\Delta$ & -0.116 3$\pm$0.01 \\
$L$           & 0.488$\pm$0.004 & $L$ & 0.488 $\pm$0.006& $L$ & 0.488$\pm$0.005 \\
$R_1$ & 0.0069$\pm$0.0004 & $R_1$ & 0.0103$\pm$0.0004 & $R_1$ & 0.0069$\pm$0.0005 \\
$R_{n>1}$ & 0.0151$\pm$0.0007 & $R_{n>1}$ & 0.0156$\pm$0.0008 & $R_{n>1}$ & 0.0153$\pm$0.0007 \\
$\delta$ & 0.935 & \multirow{2}{*}{ $\beta_1$ }& \multirow{2}{*}{  20 } & \multirow{2}{*}{$\theta_{1,\text{cr}}$} & \multirow{2}{*}{ 0.49$\pm$0.02} \\
$E_1$ & -1.9 & & & & \\
$\Delta E_2$ & -1.29$\pm$0.5 & $\beta_2$ & 9.9$\pm$5 & $\theta_{2,\text{cr}}$ & 0.66$\pm$0.1 \\
$\Delta E_N$ & 8.7$\pm$2 & $\beta_3$ & 10.2$\pm$4 & $\theta_{\infty,\text{cr}}$ & 0.0 \\
$N_0$ & 10  & $\alpha$ & 1.13$\pm$0.4 & $N_0$ & 2.86$\pm$0.6 \\
\hline

\end{tabular}
\end{center}
\caption{\label{tab:dip_pars}Fit parameters for the three fits shown in Fig.~\ref{fig:dip_xray}}  
\end{table*}

An interesting feature of Fig.~\ref{fig:dip_xray}C, in both the AFM data and the models, is that the rms roughness at film thickness $\Theta=2$ is \textit{smaller} than that at $\Theta=1$. This is reflected in the fact (see Table~\ref{tab:dip_pars}) that the parameters controlling interlayer transport for all three models do not change monotonically, as in Table~\ref{tab:pent_pars}. The cause and meaning of this behavior are outside the scope of this paper: however we suggest that it could be related to the structural evolution of the DIP film~\cite{Kowarik:2009og} or the dependence of growth rate on coverage. 

The key difference between Tables~\ref{tab:dip_pars} and~\ref{tab:pent_pars} is the additional parameter,  R$_{n>1}$, corresponding to the late-time growth rate of the film. Above, we found that the AFM data in Fig.~\ref{fig:afm_thick} gave a late-time growth rate of 0.0156 ML/s and a growth rate acceleration of 111\%.  Referring to Tab.~\ref{tab:dip_pars}, we find that this late-time growth rate agrees well with that obtained from all three models. The growth rate acceleration for the mC and FZ models yields 112\% and 122\%, respectively, in excellent agreement with the result from AFM. The BK model, which we reiterate is not intended to model growth rate acceleration accurately, gives a more modest growth rate acceleration of 56\%. We conclude that, at least for systems exhibiting relatively smooth growth at early times, the mC and FZ models provide  an accurate means of determining growth rate acceleration. 

\section{Summary and Conclusions}

In this paper, we have provided the first comparative study of the problem of obtaining surface morphology from \insitu, time resolved XRR data. We described three mean-field models of thin film growth, along with detailed examples of their application to quantitative analysis of XRR data obtained at the anti-Bragg position during growth. Two sets of organic thin films were grown using hyperthermal deposition, pentacene/SiO$_2$ and DIP/SiO$_2$. For each system, \insitu XRR data was obtained during growth of four films, grown to different thicknesses, under nominally identical conditions. The XRR data were fit to each of the the three models, resulting in detailed simulations of the time-dependent morphology of each film. Finally, these simulated morphologies were directly compared with AFM data from each of the four films to critically evaluate the quality of each simulation.

We find, first, that all of the models provide good descriptions of both the XRR data and, at early times, the surface morphology of an evolving film. This is significant, since the evolution of layer coverages incorporates the rate of interlayer transport, which controls the onset of roughness. Second, we find that the model which fits a particular data set best statistically, i.e. which gives the lowest value of $\chi_\nu^2$, does not necessarily provide the most faithful reproduction of the true surface morphology. The models we used all incorporate analogous parameters determining interlayer transport as a function of layer coverage, but these parameters represent fine distinctions regarding intralayer atomistic kinetics and morphology. The fact that better fits do not correlate to improved agreement with the true surface morphology suggest that our measurements are not sensitive to such fine distinctions.

We understand this lack of sensitivity to fine detail as resulting from the limited amount of information contained in the intensity at a single point $q_z$  on the reflectivity curve. As stated above, \insitu studies performed at multiple $q_z$ values have been demonstrated, both in angular dispersive\cite{Kowarik:2009ip} and energy dispersive modes using white beam\cite{Kowarik:2007kf}. However, these techniques both suffer from poorer time resolution than in single point mode. In angular dispersive mode, this results from the requirement of moving the sample and/or sample chamber during the growth. In energy-dispersive mode, the time resolution is limited by detector dynamic range\cite{Kowarik:2007kf}, and the fact that the signal is so much stronger near Bragg peaks compared to the anti-Bragg position. A variation to this approach would be to combine an energy-dispersive detector with a finite bandwidth beam, for example 10\%, obtained with wide bandwidth multilayers\cite{Kazimirov:2006ip} in conjunction with a bend magnet or wiggler. Using a 10$\pm$0.5 keV beam, an energy resolution of 0.2 keV would result in approximately 4-5 independent points on the reflectivity curve being obtained simultaneously, resulting in significantly more information with minimal loss in time resolution.

Concerning the determination of growth rate: in the case of pentacene/SiO$_2$, which exhibits an abrupt roughening transition near 2 ML, different models obtain values for the total growth rate that vary by as much as 10\%. For DIP/SiO$_2$, which exhibit extended LBL oscillations, we find that fits of XRR data to the mC and FZ models both correctly extract the degree of growth rate acceleration. 

A key feature of the mC and FZ models is that they both incorporate measurable quantities, such as the layer step-edge density and the critical coverage for next-layer nucleation, as fit parameters. Future work will aim to test whether these parameters, rather than only the layer coverages, can be extracted from XRR data during growth. For example, we will investigate how accurately the critical coverage parameter, $\theta_{\text{cr}}$ extracted from fits of XRR data to the FZ model, corresponds to independent measurements of $\theta_{\text{cr}}$. We believe that such work will contribute significantly to the technologically important challenge of understanding, and ultimately controlling, physical and experimental factors determining surface morphology in thin films. 

\appendix

\section{Thin Film Scattering Amplitudes Parameterized for \Insitu XRR } \label{appendixa}
Here, we derive Eq.~\ref{eq:xray2} from Eq.~\ref{eq:xray1}, resulting in approximate values of $\Afilm/\Asub$ and  $\phisub$ for the technologically relevant case of an organic thin film on a substrate with a thin interfacial layer. Referring to Fig.~\ref{fig:real_schem}, we begin by dividing the integral in Eq.~\ref{eq:xray1} into separate volumes, treating the buried substrate, the interfacial layer, and each crystalline layer of the film as distinct. Since we are only interested in the specular intensity, we ignore the fact that the film layers are only partially complete, and instead treat the coverage $\theta_n$ as a modification to the density $\rho_2$. This step ignores the $q_{\|}$ contribution of the diffuse scattering\cite{Sinha:1996tw,Ferguson:2009ig}, which depends on both details of the system as well as instrumental resolution. Neglecting the static roughness of each interface, (which can be easily incorporated\cite{Krause:2004ss} if desired), and allowing only $\theta_1$ to be nonzero, Eq.~\ref{eq:xray1} becomes
\begin{multline}
  \label{eq:xray3}
A(q_z) /A_0  = \rho_0 \int_{-\infty}^{-T_1} dz e^{-iq_zz}  + \rho_1  \int_{-T_1}^0 dz e^{-iq_zz} \\
 +\theta_1 \int_0^{c+2\Delta_z} dz  \rho_2(z) e^{-iq_zz} \text{,} 
\end{multline}
where $A_0$ is the sample area, and $2\Delta_z$ is a correction term for the height of the first layer.

We next carry out the integrals, also allowing layers $n>1$ to be nonzero:
\begin{multline}
  \label{eq:xray4}
A(q_z) /A_0 = \frac{i}{q_z} \left [ \rho_1 + (\rho_0-\rho_1)e^{i q_z T_1}  ) \right ] \\ 
+ \Amol(q_z) e^{-iq_z (c/2+\Delta_z)}  \sum_{n=1}^N \theta_n e^{-i (n-1) q_z c} \text{.} 
\end{multline}
The term $\Amol(q_z)$ is the molecular structure factor, defined as
\begin{equation}
  \label{eq:xray5}
\Amol(q_z) = \sum_m e^{-i q_z \zeta_m} \int dz \rho_m(z) e^{-iq_z z}
\text{,} 
\end{equation}
where the sum is over atoms in a unit cell, and $\zeta_m$ is the height of each atom relative to the center of the molecule (defined as $z=c/2+\Delta_z$) above the substrate, and $\rho_m(z)$ is the linearly-projected electron density of atom $m$. Finally, we factor out the term $e^{-iq_z (c/2+\Delta_z)}$  in Eq.~\ref{eq:xray4}, and convert to reciprocal lattice units $L=q_z/(2\pi/c)$, defining $\tau\equiv T_1/c$, $\Delta\equiv \Delta_z/c$. Equation \ref{eq:xray4} then becomes
\begin{multline}
\label{eq:xray6}
A(q_z) /A_0 = \frac{c}{2\pi L} i e^{i \pi L (1+2\Delta)}(\rho_1 + (\rho_0-\rho_1)e^{i 2\pi L\tau} ) \\+ \Amol(q_z) 
\sum_{n=1}^N \theta_n e^{-i (n-1) 2\pi L}\text{,}
\end{multline}
in which the first term is equivalent to $\Asub e^\phisub$ in Eq.~\ref{eq:xray2}. Equation~\ref{eq:xray6} thus accomplishes the goal of explicitly obtaining the parameters $\phisub$ and $\Afilm/\Asub$ in Eq.~\ref{eq:xray2} in terms of physical parameters of the film.  We note too, that for centrosymmetric molecules such as pentacene, $\Amol$ is pure real, so that at the anti-Bragg position ($L=0.5$), the second term contributes only pure real terms to the complex sum. For non-centrosymmetric molecules, $\Delta_z$ may still be chosen such that $\Amol(q_z)$ is pure real, in which case $z=c/2+\Delta_z$ may not correspond to the molecular center. Finally, we note that Eq.~\ref{eq:xray6}, with explicit computation of $\Amol(q_z)$ from Eq.\ref{eq:xray5} can be used to calculate the entire specular reflectivity, including Bragg peaks, for reasonably thin films, e.g. $T_{\text{film}} \ll$ 1$\mu$m.

For large, organic molecules, the anti-Bragg position probes density fluctuations on an approximate length scale $l=2\pi/q_z=c/L$,  which is large compared to interatomic distances. As a result, a molecular layer is well approximated by a uniform density slab. In this approximation, $\Amol$ can be written
\begin{equation}
  \label{eq:xray7}
\Amol(q_z) = 2 \rho_2 \frac{c}{2\pi L} \sin(\pi L)
\text{.} 
\end{equation}
We note that the simple case of homoepitaxy can be recovered by setting the densities equal and $\tau=0$. The uniform slab approximation at the anti-Bragg position $L=0.5$ then gives the familiar result\cite{Kowarik:2009ie}:
\begin{equation}
I_{\text{AB}} \propto \left | A_{\text{AB}} \right |^2=\left | 1-2\sum_{n=1}^N \theta_n (-1)^{n-1}   \right |^2
\end{equation}

\section{Scattering Amplitude Evolution During LBL and SF Growth on Vicinal Surfaces.} \label{appendixb}

To compare trajectories of the total scattering amplitude during LBL and SF growth, we imagine an ideal, vicinal substrate with step-height $c$ and terrace width $W$, and with the positive $x$ direction perpendicular to, and pointing towards uphill steps. We choose the origin to coincide with the center of a terrace in $x$ and height $z$ of the first growing layer. The scattering amplitude $A_\vic$ of a film with a total of $N_L$ partially complete layers (compare with Eq.~\ref{eq:xray2}) is then 
\begin{multline}\label{eq:sf1}
A_\vic  =\sum_{k=-N_W}^{N_W} e^{-i (q_z c + q_x W)k } \times \\ \left [ \Asub e^{i \phisub} + \sum_{n=1}^{N_L} e^{-iq_zc(n-1)}A_{\ter}(\theta_n) \right ]\text{,}
\end{multline}
where $A_{\ter,n}(\theta_n)$ is the scattering amplitude of layer $n$ with fractional coverage $\theta_n$. The left-hand sum in Eq.~\ref{eq:sf1} is over $N_T=2 N_W+1$ terraces, and defines the specular condition as $q_z c=-q_x W$. In perfect SF growth, each layer $n$ commences only when layer $n-1$ is complete, and grows from the up-hill step at $x=W/2$, advancing from right to left until reaching the down-hill step at $x=-W/2$ at $\theta_n=1$. Thus, 
\begin{equation} \label{eq:sf2}
A_{\ter,\sflow}(\theta_n)  = f \int_{W/2-\theta_n W}^{W/2}dx e^{-iq_x x},
\end{equation}
where $f$ is the electron density per unit length of the growing layer.  Along the specular rod, we can substitute $-q_zc$ with $q_xW$ in all of the terms $e^{-iq_zc(n-1)}$ in Eq.~\ref{eq:sf1}. Next we can bring these prefactors into the integrands and substitute variables, so that the right-hand term in brackets becomes
\begin{multline} \label{eq:sf3}
\sum_{n=1}^{N_L} e^{-iq_zc(n-1)}A_{\ter}(\theta_n) =  f \sum_{n=0}^{N_L-2} \int_{-W/2-nW}^{W/2-nW}dx e^{-iq_x x} \\
+  f \int_{W/2- (N_L-1 + \theta_{N_L})W}^{W/2 - (N_L-1)W} dx e^{-iq_x x} 
\end{multline}
All of the integrals in Eq.~\ref{eq:sf3} may now be combined. Identifying the film thickness $\Theta = N_L-1 + \theta_{N_L}$, we have
\begin{multline}
\label{eq:sf4}
\sum_{n=1}^{N_L} e^{-iq_zc(n-1)}A_{\ter,n}(\theta_n)  =  f \int_{W/2-W\Theta}^{W/2}dx e^{-iq_x x} \\
 = \frac{f}{iq_x}e^{-iq_x W/2} \left [ e^{iq_x W \Theta}   - 1 \right ].
\end{multline}
for the total film scattering amplitude along the specular rod. Eq.~\ref{eq:sf4} shows that the scattering amplitude in SF growth completes a circle in the complex plane whenever $q_xW\Theta = 2\pi n$, or equivalently (recalling $-q_xW=q_zc = 2\pi L$), whenever $\Theta = n/L$.

In LBL growth,  deposited material lands randomly on each terrace, so the scattering amplitude of a partially filled layer is just Eq.~\ref{eq:sf2} \textit{evaluated} at $\theta=1$, then multiplied by the actual coverage $\theta_n$: 
\begin{equation} \label{eq:sf5}
A_{\ter,\lbl}(\theta_n)  = \theta _n fW\frac{\sin(q_x W/2)}{q_x W/2}.
\end{equation}
Just like Eq.~\ref{eq:xray2}, Eq.~\ref{eq:sf1} with the substitution of Eq.~\ref{eq:sf5} results in regular, polygonal trajectories in the total scattering amplitude. 

The effect of considering a vicinal rather than singular substrate is to reduce the scattering amplitude in the specular direction by a factor $\sin(q_x W/2)/(q_x W/2)$. At the anti-Bragg position, $q_xW=\pi$ for a vicinal surface, but $q_x=0$ for a singular surface, implying that the scattering amplitude of a vicinal terrace is $2/\pi$ that of the equivalent area on a singular surface. This decrease in intensity from the specular, anti-Bragg position is redistributed into off-specular angles due to diffraction from the steps, which form a blazed grating. Scattered intensity will appear at $q_xW = -\pi+ 2 \pi n$ for all integers $n$. The total scattered intensity in the plane $q_z = \pi/c$ can be summed by adding up the square of Eq.~\ref{eq:sf5} over all allowed $q_x$. Noting the identity\cite[Eq. 1.422.4, p. 44]{Gradshtein:2007},
\begin{multline}
\sum_{n=-\infty}^\infty \left ( \frac{\sin(x + \pi n)}{x + \pi n} \right )^2 = \sin^2 x \sum_{n=-\infty}^\infty \frac{1}{(x+\pi n)^2} = 1,
\end{multline}
we find that the total scattered intensity in the plane $q_z = \pi/c$ for both vicinal and singular surfaces is $(\theta_n f W)^2$. This is actually an example of a more general sum rule easily derived from Ref.~[\onlinecite{Sinha:1988qb}]: the total scattered intensity at a given $q_z$ is independent of the surface morphology, provided that the lateral length scale of roughness is small compared to the coherence length.

In homoepitaxy, the circular trajectory in Eq.~\ref{eq:sf4} for the scattering amplitude in the specular condition must result in constant intensity. In that case, and making use of Eq.~\ref{eq:sf5}, the term $\Asub e^{i \phisub}$ is evaluated as :
\begin{equation} \label{eq:sf6}
\begin{split}
\Asub e^{i \phisub} & = \lim_{\epsilon \rightarrow 0}A_\ter \sum_{n=1}^\infty e^{iq_z c n - \epsilon n} \\
& =  A_\ter \frac{ e^{iq_z c}}{1-e^{iq_z c}} \\
& = 2f \frac{\sin(q_x W/2)}{q_x} \frac{ e^{iq_zc/2}}{-2i \sin(q_zc/2)} \\
& = \frac{f}{iq_x}{e^{-iq_xW/2}},
\end{split}
\end{equation}
where $A_\ter$ is Eq.~\ref{eq:sf5} evaluated at $\theta_n=1$, and we have again used $q_zc=-q_xW$. Adding Eq.~\ref{eq:sf6} to Eq.~\ref{eq:sf4} translates the circle to the origin, so that the scattered intensity is constant with increasing $\Theta$.

\begin{acknowledgements}

This work was supported by the Cornell Center for Materials Research, a National Science Foundation Materials Research Science and Engineering Center (NSF-DMR-0520404), and was performed in part at the Cornell High Energy Synchrotron Source, also supported by the National Science Foundation and NIH-NIGMS (NSF-DMR-0225180).  JRE acknowledges supplementary support via NSF-ECS-0210693 and NSF-ECS-0304483. 

\end{acknowledgements}


\end{document}